\documentclass{emulateapj}

\newcommand{\eps}[1]{\mbox{log~$\epsilon$(#1)}} 
\newcommand\iso[2]{$^{\rm #1}$#2}
\newcommand{\nion}[2]{#1\,\lowercase{{\sc\thinspace #2}}}
\newcommand\wave[1]{\mbox{$\lambda$#1\,\AA}}
\def\bd{BD+17\deg 3248}
\def\cs{\mbox{CS~22892-052}}
\def\deg{{$^{\circ}$}}
\def\eg{\mbox{\it e.g.}}
\def\etal{\mbox{et al.}}
\def\hd{\mbox{HD~221170}}
\def\ie{\mbox{\it i.e.}}
\def\invcm{\mbox{cm$^{\rm -1}$}}
\def\kmsec{\mbox{km~s$^{\rm -1}$}}
\def\logg{\mbox{log~{\it g}}}
\def\loggf{\mbox{log~{\it gf}-value}}

\def\teff{\mbox{$T_{\rm eff}$}}
\def\third{{3$^{\rm rd}$}}
\def\vmicro{\mbox{$\xi_{\rm t}$}}

\def\Yu{\mbox{\citet{yu+05}}}

\shorttitle{Chemical Composition of HD~221170}
\shortauthors{Ivans et al.}

\begin{document}

\title{
	Near-UV Observations of HD~221170: \\ New Insights into the Nature of $r$-Process-Rich Stars\footnote{Some of the data presented herein were obtained at the W.\ M.\ Keck Observatory, which is operated as a scientific partnership among the California Institute of Technology, University of California, and NASA, and was made possible by the financial support of the W.\ M.\ Keck Foundation. This paper includes data taken at the McDonald Observatory of The University of Texas at Austin.}}

\author{
Inese I.\ Ivans\altaffilmark{2,3,4}, 
Jennifer Simmerer\altaffilmark{5},
Christopher Sneden\altaffilmark{5},
James E.\ Lawler\altaffilmark{6},
John J.\ Cowan\altaffilmark{7},
Roberto Gallino\altaffilmark{8,9}, and
Sara Bisterzo\altaffilmark{8}
}
\altaffiltext{2}{The Observatories of the Carnegie Institution of 
	Washington, 813 Santa Barbara St., Pasadena, CA 91101; 
	iii@ociw.edu}

\altaffiltext{3}{Princeton University Observatory, Peyton Hall, Princeton, 
	NJ 08544}

\altaffiltext{4}{Carnegie-Princeton Fellow}

\altaffiltext{5}{Dept.\ of Astronomy, The University of Texas, 
	Austin, TX 78712; jensim@astro.as.utexas.edu,
	chris@verdi.as.utexas.edu}

\altaffiltext{6}{Dept.\ of Physics, University of Wisconsin, Madison, WI 
	53706; jelawler@.wisc.edu}

\altaffiltext{7}{Homer L.\ Dodge Dept.\ of Physics and Astronomy, University of Oklahoma, 
	Norman, OK 73019; cowan@nhn.ou.edu}

\altaffiltext{8}{Dipartimento di Fisica Generale, Universita' di Torino,
        Via P.\ Giuria 1, 10125 Torino, Italy; gallino@ph.unito.it, 
	bisterzo@ph.unito.it}

\altaffiltext{9}{Centre for Stellar and Planetary Astrophysics, 
        School of Mathematical Sciences, P.O.\ Box 28M, Monash 
	University, Victoria 3800 Australia}

\begin{abstract}
Employing high resolution spectra obtained with the near-UV sensitive 
detector on the Keck~I HIRES, supplemented by data obtained with the 
McDonald Observatory 2-d coud\'e, we have performed a comprehensive 
chemical composition analysis of the bright $r$-process-rich metal-poor 
red giant star \hd.  Analysis of 57 individual neutral and ionized 
species yielded abundances for a total of 46 elements and significant 
upper limits for an additional five.  Model stellar atmosphere 
parameters were derived with the aid of $\sim$200 Fe-peak transitions.  
From more than 350 transitions of 35 neutron-capture ($Z >$ 30) species, 
abundances for 30 neutron-capture elements and upper limits for three 
others were derived.  Utilizing 36 transitions of La, 16 of Eu, and 
seven of Th, we derive ratios of \eps{Th/La} = $-$0.73 ($\sigma$ = 0.06) 
and \eps{Th/Eu} = $-$0.60 ($\sigma$ = 0.05), values in excellent 
agreement with those previously derived for other $r$-process-rich 
metal-poor stars such as \cs, \bd, and HD~115444.  Based upon the Th/Eu 
chronometer, the inferred age is 11.7 $\pm$ 2.8~Gyr.  The abundance 
distribution of the heavier neutron-capture elements ($Z \ge$ 56) is fit 
well by the predicted scaled solar system $r$-process abundances, as 
also seen in other $r$-process-rich stars.  Unlike other 
$r$-process-rich stars, however, we find that the abundances of the 
lighter neutron-capture elements (37 $< Z <$ 56) in \hd\ are also 
statistically in better agreement with the abundances predicted for the 
scaled solar $r$-process pattern.

\end{abstract}

\keywords{nuclear reactions, nucleosynthesis, abundances -- 
Galaxy: evolution -- 
Galaxy: abundances -- 
stars: abundances -- 
stars: Population II -- 
stars: individual (\objectname{HD~221170}) 
}

\section{INTRODUCTION\label{intro}}

The bulk of the neutron-capture ($n$-capture) elements beyond the iron 
group are created by some combination of the slow and rapid 
neutron-capture nucleosynthesis processes ($s$- and $r$-process), with 
each responsible for approximately half of the isotopes.  In the 
$r$-process, both the neutron density and neutron flux are high.  
Neutron-rich sites associated with massive star core collapse supernovae 
(SNeII) are the likeliest sites for the $r$-process (see \eg, Cowan \& 
Thielemann, 2004\nocite{ct04} and references therein), although the 
astrophysical site of the $r$-process has yet to be identified.  
Possible sites for the $r$-process include: supernovae winds/hot bubbles 
(see \eg, Hoffman, Woosley, \& Qian 1997\nocite{hwq97}; Terasawa \etal\ 
2002\nocite{terasawa+02}; Wanajo \etal\ 2002\nocite{wanajo+02}; Kohri, 
Narayan, \& Piran 2005\nocite{knp05}; and references therein); disks and 
jets (\eg, Cameron 2003\nocite{cameron03} and references therein); 
neutron star mergers (\eg, Freiburghaus, Rosswog, \& Thielemann 
1999\nocite{frt99}; Rosswog \etal\ 1999\nocite{rosswog+99}; and 
references therein but also see Argast \etal\ 2004\nocite{argast+04} for 
a contrasting view) and/or neutron star formation during accretion 
induced collapse (\eg, Wheeler, Cowan \& Hillebrant 1998\nocite{wch98};
Cohen et al.\ 2003\nocite{cohen+03}; Qian \& Wasserburg 
2003\nocite{qw03}). 

Among the isotopes formed in the $r$-process are the radioactive group 
of elements known as the actinides, which include isotopes of Th and U.  
Due to their known radioactive decay rates, the abundances of Th (and 
U) in low-metallicity stars have been employed to derive the ages of 
presumably some of the oldest stars in the Galaxy, thereby setting a 
minimum for the age of the Universe.  Critical assumptions in the
analysis of the observations are that the production ratios of the 
elements are known, and that the elements under investigation arise
from the same nucleosynthetic site.  Following earlier work on the
derivation of Th abundances and/or Th-based ages by \citet{butcher87},
\citet{pagel89}, and Fran\c{c}ois, Spite \& Spite (1993\nocite{fss93}), 
\citet{sneden+96} derived the first Th/Eu-based nucleocosmochronometric 
age for a very metal-poor star: CS~22892-052, whose extreme $r$-process
abundance enhancements were discovered by \citet{mcw+95}. This was 
followed soon after by other Th/Eu-based age determinations of very 
metal-poor stars by \citet{pkt97}, \citet{cowan+99,cowan+02}, 
\citet{sneden+00b}, \citet{jb01}, and \citet{hill+02}.  In most of 
these studies, the use of Th/Eu cosmochronometry has yielded consistent 
results (\eg, \cs, HD~115444, and \bd).  A weak U detection was made in 
\bd\ \citep{cowan+02} and an upper limit was derived for \cs\ 
\citep{sneden+00a,sneden+03}, permitting lower limits to be placed on 
ages inferred by the Th/U ratio, which turn out to be in accord with 
those derived from Th/Eu cosmochronology.  However, in the case of 
CS~31082-001, for which the first U abundance determination was made in 
an ultra metal-poor star \citep{cayrel+01}, the age inferred from Th/Eu 
is found to be significantly different from that inferred from Th/U 
\citep{hill+02}. Discrepant age results have also been reported for
HD~221170 by \Yu\ employing the abundances they derived for Th/U and 
Th/Eu.

The fundamental assumption built into the technique of applying these
abundances to derive ages via nucleocosmochemistry is that the elements 
are created in the same processes, \ie\ the same nucleosynthetic sites, 
and track the contributions of that process through Galactic chemical 
evolution, \ie\ a production ratio can be specified.  There is no doubt 
that all of the Th and U is produced in the $r$-process: the 
termination point of the $s$-process occurs at $^{209}$Bi.  Isotopes 
heavier than $^{209}$Bi decay too quickly to be built by the 
$s$-process.  And, Eu is predominantly an $r$-process element -- over 
90~\% of the solar Eu was produced this way 
\citep{ag89,kbw89,arlandini+99,burris+00,jensim+04,travaglio+04}.  However, 
Th lies 29 atomic mass units away from Eu.  It has been argued that the 
production of the actinide elements may not be known sufficiently well 
to trust their use as nucleocosmochronometers (see \eg, Arnould \& 
Goriely 2001\nocite{ag01}, and references therein).  Certainly, with 
lighter $n$-capture isotopes, there is evidence that multiple 
$r$-process sites or sources have contributed to the total abundance in 
both the Sun and in metal-poor stars.  While at low metallicities the 
observed abundances of $r$-process-rich stars in the atomic range of 
$Z \ge$ 56 are in good agreement with the predicted scaled solar 
$r$-process patterns \citep{sneden+96,sneden+98,cowan+99}, the 
abundances of the lighter $n$-capture elements are not 
\citep{sneden+00a}.  Furthermore, based on studies of the inferred 
abundances of short-lived isotopes in the early solar system, multiple 
sites or sources have been required to explain the early solar system 
abundances of both light and heavy isotopes normally considered to be 
of primary $r$-process origin (see \eg, Cameron, Thielemann, \& Cowan 
1993\nocite{ctc93}; Wasserburg, Busso, \& Gallino 1996\nocite{wbg96}; 
Meyer \& Clayton 2000\nocite{mc00}).

In the Sun, the abundances are the integrated result of many 
generations of stars, including millions of SNeII, and depend upon the 
details of the star formation history, initial mass function, chemical 
yields, etc.  However, the heavy element abundances most useful for 
unravelling the origins of the $r$-process correspond to those 
observed in stars which formed from material with little prior 
nucleosynthetic processing, such as the relatively pristine material 
out of which extremely metal-poor stars were born.  To further explore 
the $n$-capture elemental abundances in $r$-process-rich stars, we 
observed the bright ($V$ = 7.7) metal-poor red giant branch star \hd.

HD~221170 has been the subject of over forty years of spectroscopic 
studies (e.g., see the comprehensive listing compiled by Gopka et al.\ 
2004\nocite{gopka+04}; their Table 1).  As first noted by \cite{wall+63}, 
the star is metal-deficient.  Recent estimates of the metallicity range 
$-$2.20 $<$ [Fe/H]\footnote{\footnotesize 
We adopt the usual spectroscopic notation that for elements A and B, 
\eps{A} $\equiv$ {\rm log}$_{\rm 10}$(N$_{\rm A}$/N$_{\rm H}$) + 12.0, 
and [A/B] $\equiv$ 
{\rm log}$_{\rm 10}$(N$_{\rm A}$/N$_{\rm B}$)$_{\star}$ $-$
{\rm log}$_{\rm 10}$(N$_{\rm A}$/N$_{\rm B}$)$_{\odot}$. e.g., 
(N$_{\rm Ho}$/N$_{\rm Fe}$)$_{\star}$ = 
5$\times$(N$_{\rm Ho}$/N$_{\rm Fe}$)$_{\odot}$ $\Rightarrow$ [Ho/Fe] = 
$+$0.7.  Also, metallicity in our discussions refers to the normalized 
iron abundance, the stellar [Fe/H] value.} 
 $<$ $-$1.96.  Included in the $n$-capture abundance study by 
\citet{gilroy+88}, HD~221170 has long been recognized as an
$r$-process-rich star and has often been utilized as a template 
metal-poor star observation in other programs, including those of
\citet{burris+00}, \citet{fulbright00}, \citet{mk01}, 
\citet{mishenina+02}, \citet{yu+02}, \citet{jensim+04}, and 
\citet{barklem+05}.  In this paper, we describe our observations and 
analysis and compare our results with previous observations, with 
scaled solar $r$-process predictions, and with other $r$-process-rich 
stars, concluding with discussions regarding the 
nucleocosmochronometric age of HD~221170 and $r$-process sites.

\section{OBSERVATIONS AND REDUCTIONS\label{obs}}

We gathered new high resolution, high signal-to-noise (S/N) spectra of 
\hd\ with the High Resolution Echelle Spectrometer (HIRES; Vogt et al.\ 
1994\nocite{vogt+94}) on the Keck~I telescope at the W.\ M.\ Keck 
Observatory and with the ``2d-coud\'e'' \'echelle spectrograph 
\citep{tull+95} on the 2.7-m H.\ J.\ Smith telescope at McDonald 
Observatory.

We observed \hd\ with HIRES at Keck~I using a blue configuration (HIRESb) 
and the new 3-chip mosaic of MIT-LL CCDs.  Using the same setup as that 
described in \citet{ivans+05}, we obtained essentially continuous 
wavelength coverage in the range $\sim$3050 $\le \lambda \le$ 5895~\AA\ 
and a resolving power of $R$ $\equiv$ $\lambda$/$\Delta\lambda$ $\simeq$ 
40,000.  Spectra of the hot rapidly rotating star $\delta$ Cet aided in 
the division of telluric features in the reddest wavelengths, and also 
served as a check on the data reduction of the bluest orders of the 
spectrum of \hd.  Four exposures of \hd\ were taken to attain a co-added 
S/N of 85:1 per resolution element at \wave{3200}, increasing redwards, 
to $\sim$120 at \wave{3520}, $\sim$140 at \wave{3900}, $\sim$260 at 
\wave{5100}, and $\sim$290 at \wave{5900}.  Data reduction was performed 
using standard tasks in IRAF\footnote{\footnotesize IRAF is distributed 
by NOAO, which is operated by AURA, under cooperative agreement with the 
NSF.} including bias subtraction, bad pixel interpolation, wavelength 
calibration, and co-addition of the one-dimensional spectra;  in
FIGARO\footnote{\footnotesize FIGARO is provided by the Starlink Project 
which is run by CCLRC on behalf of PPARC (UK).} including flat fielding, 
light cosmic ray excision, sky and scattered light subtraction, and 
extraction of the one-dimensional spectra; and in SPECTRE \citep{fs87} 
for final processing including continuum normalization and telluric 
feature division.

As part of the \citet{jensim+04} survey of La and Eu abundances over a
large metallicity range, data for \hd\ were also gathered with the 
McDonald Observatory 2d-coud\'e \'echelle spectrograph in the wavelength 
range 3800~\AA\ $< \lambda <$ 7800~\AA.  The spectrum is continuous in 
the range $\lambda <$ 5800~\AA, with some losses to order interstices 
at redder wavelengths.  We also acquired a spectrum of the hot, rapidly 
rotating star $\zeta$~Aql for use in cancelling the yellow and red 
telluric features of O$_{2}$ and H$_{2}$O.  The spectrograph set-up 
yielded $R$ $\simeq$ 60,000.  Data reduction 
was performed using standard tasks in IRAF and SPECTRE.  We refer the 
reader to Simmerer et al.\ (their \S~2\nocite{jensim+04}) for further 
details regarding the spectrograph set-up and data reduction of the 
2d-coud\'e observations.  The co-added reduced 2d-coud\'e spectra have 
S/N $>$ 260 for $\lambda >$ 5900~\AA, and this declines steadily to 
levels of 80 at $\lambda$ = 4000~\AA. 

\begin{figure}
\epsscale{0.9}
\plotone{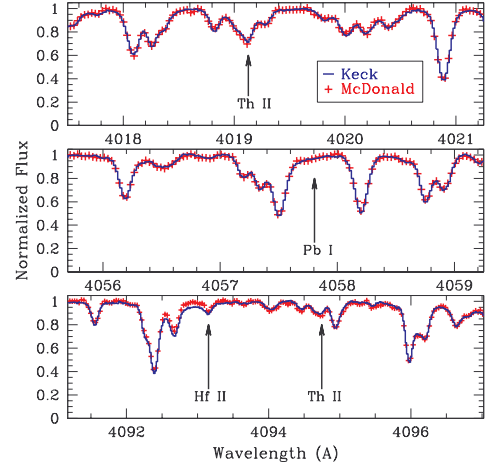}
\caption{
\label{fig.spectra} 
Examples of the reduced spectra taken with the Keck HIRESb (blue 
histogram) and McDonald 2d-coud\'e spectrographs (red crosses) for a 
sample of the wavelength regions displayed by Yushchenko et al.\ (2002, 
their Figure 2; and 2005, their Figures 1--4): \nion{Th}{ii} at 
4019.13~\AA; \nion{Pb}{i} at 4057.81~\AA; \nion{Hf}{ii} at 4093.16~\AA;  
and \nion{Th}{ii} at 4094.75~\AA.
}
\end{figure}

In Figure~\ref{fig.spectra}, we display selected spectrum swaths for our 
data sets in overlapping spectral regions.  Two of us independently 
reduced the Keck and McDonald spectra, employing different methods and 
completely different software for the removal of scattered light and 
cosmic ray contributions.  The resulting spectra are in excellent 
agreement.  The selected features in this figure are the same as those 
displayed by Yushchenko et al.\ (2002\nocite{yu+02}, their Figure 2; 
and 2005\nocite{yu+05}, their Figures 1--4).  Their 2005 investigation 
is based on high S/N, $R$ $\simeq$ 45,000 data gathered at the 2.0-m 
Zeiss telescope at the Peak Terskol Observatory with the coud\'e 
\'echelle spectrometer \citep{musaev+99}, and should be comparable to 
our Keck spectrum.  However, the appearance of our spectra do not 
compare favourably with those displayed in the Yushchenko et al.\ 
study.  Neither do our Keck and McDonald data sets match well the 
spectrum displayed in the study by \citet{gopka+04}, who based their 
investigation on previously acquired Peak Terskol data as well as high 
S/N, $R$ $\simeq$ 60,000 data gathered with the ELODIE \'echelle 
spectrograph \citep{skc98} on the 1.9-m telescope of the Observatoire 
de Haute Provence.  In addition, Figure 1 of \citet{gopka+04} shows 
that the Peak Terskol and ELODIE data appear to be in less than good 
agreement with each other.  However, all data sets in this discussion 
are purportedly of high (100--250) S/N; the cause of the mismatches is 
unclear.

Carney et al.\ (2003\nocite{carney+03}; their Table 4) find that \hd\ 
has possessed a constant radial velocity during the 14 years spanned
by their observing programme.  This star does not appear to have a 
companion.  The four data sets under discussion here were gathered 
over a number of years, with those of this study taken more than three 
years apart (ELODIE, prior to 1998; McDonald, July 2001; Peak Terskol, 
June 2002; and Keck, October 2004).  It is highly unlikely that the 
data belong to a variable star.   We do not have an explanation for 
either (i) the difference between the appearance of the two 
\citep{gopka+04} data sets; or (ii) the difference between the Gopka 
et al./Yushchenko et al.\ data sets and the spectra employed in this 
study.  Further comments on these differences will be given in 
\S~5.

\section{ANALYSIS\label{analysis}}

Our abundance analysis relied on the results of a combination of 
spectrum syntheses and equivalent width (EW) analyses. We relied on the 
Keck HIRESb data for wavelengths in the range $\sim$3050 
$\le \lambda \le$ 5895~\AA, and on the McDonald 2d-coud\'e data for 
redder wavelengths.  In overlapping wavelength regions, both data sets 
were checked in the analyses of particularly weak or noisy features.  
For each spectral order, the continuum was set by interactively fitting 
a spline function to line-free spectral regions. Locating the continuum 
was aided by comparing the spectra of our program stars with that of 
the Arcturus atlas \citep{griffin68}, as well as spectrum syntheses.  
The following sub-sections describe the linelists we used, the 
measurements we performed, the solar abundances we adopted, the methods 
we employed, and the stellar parameters and abundances we derived in 
this study.

\subsection{Equivalent Width Measurements\label{anal-ew}}

In Table~\ref{hd.analysis}, we list relevant data for all of the 
transitions employed in this study.  The elements are listed in order 
of atomic number ($Z$), and for each element, the values for neutral 
species precede those of the ionized ones, and are otherwise listed in 
order of increasing wavelength.  

Parameters for the individual transitions ($\lambda$, $\chi$, log~$gf$,
and EW) are presented in Columns 2--5.   The EWs were measured with 
SPECTRE, using either direct integration of the flux across an observed 
line profile, or adopting a Gaussian approximation (for all but the 
strongest lines, for which Voigt profile fits were employed).  Most of 
the EW measurements are of neutral iron lines with atomic parameters 
adopted from \citet{obrian+91}.  Supplementing these lines are features 
we have employed in other high resolution abundance studies of globular 
cluster red giant stars  (\eg, Ivans et al.\ 2001\nocite{ivans+01}; 
Sneden et al.\ 2004\nocite{sneden+04}; Johnson, Ivans \& Stetson 
2005\nocite{jis05}), $r$-process-enriched metal-poor stars such as \bd\
and \cs\ (\eg, Cowan et al.\ 2002\nocite{cowan+02}; Sneden et al.\ 
2003\nocite{sneden+03}), and other field stars with metallicities 
comparable to that of \hd\ (\eg, Ivans et al.\ 2003\nocite{ivans+03}), 
supplemented by recent laboratory results (eg., Nilsson et al.\ 
2005\nocite{nilsson+05}).  Additional notes regarding specific elements 
are discussed further in \S~\ref{anal-anal}.

\begin{figure}
\epsscale{0.9}
\plotone{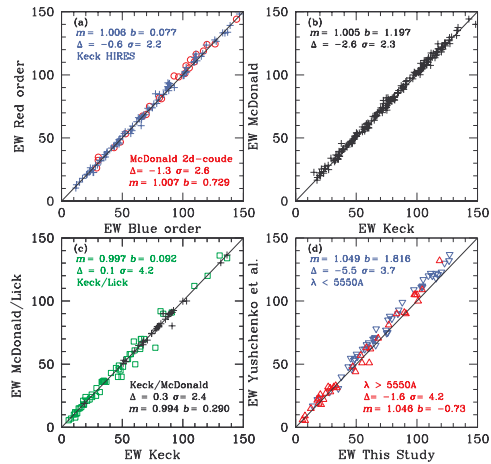}
\caption{
\label{fig.ew} 
Comparisons of EWs for lines in common between the data sets discussed 
in \S~2.  Panel (a) overlapping orders within the Keck and 
McDonald data sets (blue + and red $\bigcirc$, respectively) ; (b) 
lines in common in the Keck and McDonald data sets; (c) independently 
measured McDonald EWs (black +) and Lick EWs \citep{fulbright00} 
(green $\square$) versus Keck EWs; panel (d) lines in common with 
\citet{yu+05}, colour-coded by $\lambda <$ 5550~\AA\ (in blue 
$\bigtriangledown$) or $>$ 5550~\AA\ (in red $\bigtriangleup$).  In 
all panels, we show the values of the slope ($m$) and zero point 
shifts ($b$) of the regressions (where the solid line represents a 
one-to-one relation), along with the mean differences ($\Delta$) 
and standard deviations of the measurements ($\sigma$).  The values 
were derived employing all EWs in common; the panel displays are 
restricted to EWs $<$ 150~m\AA.
}
\end{figure}

In Figure~\ref{fig.ew}, we compare the EWs measured in the Keck HIRESb 
and McDonald 2d-coud\'e data sets. As illustrated (as well as 
quantitatively described in this figure), the data reduction and EW 
measurements in this study appear to have been performed consistently.  
For instance, in panel (a), considering the Keck and McDonald spectra
independently, we show comparisons between EWs of the same lines 
appearing on adjacent \'echelle orders that overlap in wavelength (\ie, 
the EW from the bluer order is compared against the EW obtained in the
redder order).  In panel (b), EWs of lines in common between the Keck
and McDonald spectra measured by the same person are compared.  In 
panel (c), our Keck HIRES EWs are compared with independently-measured 
McDonald 2d-coud\'e \citep{jensim+04} and Lick Hamilton 
\citep{fulbright00} EWs.  These three panels demonstrate the accuracy 
of our EWs both internally and externally, despite the numerous 
differences in spectrographs, data reduction and measurement tools, 
and techniques employed by three independent observers.  However, the 
same cannot be said of the data of panel (d), showing an EW comparison 
between this study and that of Yushchenko et al.\ (2005\nocite{yu+05}; 
their Table 1).  The EW differences are large, and also appear to be 
wavelength-dependent: they are largest in the blue wavelength regions 
where most of the $n$-capture features are to be found.  We are unable 
to explain the EW differences between \Yu\ and either (i) our study or 
(ii) an independent set of EWs measured from data acquired with the 
Lick Hamilton \'echelle spectrograph \citep{fulbright00}.  Further 
comments on these differences will be given in \S~5.

\subsection{Stellar Atmosphere Model and Parameters\label{anal-teff}}

We employed stellar atmospheres without overshooting \citep{ck04}, 
using a modified interpolation code supplied by A.\ McWilliam (2001, 
private communication).  We performed the abundance calculations with a 
current version of the LTE stellar line analysis code, MOOG 
\citep{sneden73}.  Initial stellar parameters were determined employing 
photometry from SIMBAD and the Two Micron All Sky Survey (2MASS); the 
extensions of the colour-\teff\ calibrations of \citet{aam96,aam99} by 
\citet{rm05}, using a software program supplied by I.\ Ram\'irez (2005, 
private communication); the value of E($B$$-$$V$) = 0.14 from the dust 
map calibrations of \citet{sfd98}; and a parallax of 2.30 $\pm$ 0.84, 
as measured by Hipparcos \citep{ESA97}.  

Utilizing the EW measurements presented in Table~\ref{hd.analysis}, we 
then iterated on the Fe abundances to eliminate abundance trends with 
respect to the excitation potentials (setting \teff; e.g., see Kraft
\& Ivans 2003\nocite{ki03}), EWs (setting the microturbulent velocity, 
\vmicro), and ionization state (setting \logg).  The initial \teff\ 
value of 4610~K presented abundance trends with the values of the 
excitation potentials of the lines, indicating that the photometric 
\teff\ was too warm.  This is likely a result of too high a value for 
E($B$$-$$V$); in contrast, the \citet{bh82} value is less than half 
that of Schlegel et al\nocite{sfd98}.  

Employing spectroscopic constraints, we derived final parameter values 
of \teff/\logg/\vmicro/[Fe/H] = 4510/1.00/1.8/$-$2.19.  Since the 
predicted range of E($B$$-$$V$) is so large (0.06--0.14~dex, with a 
mean of 0.1), we are satisfied with the E($B$$-$$V$) = 0.095 implied 
by the stellar parameters we derived by the spectroscopic constraints, 
which are independent of the photometry.  This reddening value, 
combined with an assumed stellar mass of 0.8~M$_{\sun}$ and the 
\logg-value we derived, yields a spectroscopically derived distance of 
675 parsecs, on the far side but within the parallax errors of the 
Hipparcos measurement.  Our derived parameters are also within the 
range of stellar parameters employed in previous CCD-based studies as 
listed in Gopka et al.\ (2004\nocite{gopka+04}; their Table 1: 4410 $<$ 
\teff\ $<$ 4686~K; 0.75 $<$ \logg\ $<$ 1.57; 1.5 $<$ \vmicro\ $<$ 
2.7~\kmsec; and $-$2.19 $<$ [Fe/H] $<$ $-$1.79).  Using spectroscopic 
constraints similar to those we employed, the \citet{gopka+04} study 
derived values of \teff/\logg/\vmicro/[Fe/H] = 4475/1.0/1.7/$-$2.03.  
\Yu\ remeasured the iron abundance, and used the parameters 
4475/1.0/1.7/$-$2.09 in their analysis, parameters in good agreement 
with the values we have derived.  

Our adopted solar abundances for each element are listed in 
Table~\ref{hd.analysis}, preceding the transition information.  With the 
exception of iron and some $n$-capture elements, our analysis relies on 
the solar photospheric (where reliable) or meteoritic abundances from 
the critical compilation of \citet{ag89}.  As in previous studies by our 
group, we adopt log~$\epsilon$(Fe) = 7.52, a value close to that 
recommended by Grevesse \& Sauval (1998; \eps{Fe} = 7.50).  We refer the 
reader to discussions by \citet{sneden+91b}; Ryan, Norris \& Beers 
(1996)\nocite{rnb96}; and \citet{mcw97}, where some of the alternative 
solar iron abundance choices are summarized.  In the element range of 57 
$< Z <$ 67, the solar abundances for six of the elements have been 
recently redetermined, employing new and improved laboratory measurements 
of atomic parameters, and we adopt those solar abundances in this study
(La -- Lawler et al.\ 2001a\nocite{lawler+01a}; 
Nd -- Den~Hartog et al.\ 2003\nocite{denhartog+03}; 
Sm -- Lawler et al. 2005\nocite{lawler+05}; 
Eu -- Lawler et al.\ 2001c\nocite{lawler+01c}; 
Tb -- Lawler et al.\ 2001b\nocite{lawler+01b}, 
and Ho -- Lawler et al.\ 2004\nocite{lawler+04}).

\subsection{Abundance Analysis Methodology\label{anal-anal}}

\hd\ is a cool star with a strong-lined spectrum.  As such, it has 
revealed many more $n$-capture transitions than we have previously 
worked with in studies of $r$-process enriched metal-poor stars such as 
CS~22892-052 \citep{sneden+03}, BD+17~3248 \citep{cowan+02}, and 
HD~115444 \citep{westin+00}.  For many elements in this study, we have 
significantly expanded the number of transitions employed in our 
analysis.  However, we have endeavoured to employ single-source and 
recent {\it gf}-values wherever possible in order to diminish the 
uncertainties involved by combining studies that may not be on the 
same {\it gf}-value system.  Happily, many of the $n$-capture element 
species detectable in metal-poor stars have been subjected to extensive 
laboratory investigations within the past two decades.  For most 
elements considered here, we have only employed {\it gf}-values 
determined in these recent lab efforts.  The exceptions will be noted 
below in \S~\ref{anal-comments}.  Laboratory transition probabilities 
were adopted without change for the $n$-capture lines of interest in 
each line list.

Abundances were derived from EW measurements for those transitions that 
we judged to be unblended and able to be modeled as single spectral 
features.  The exception to this rule was in our treatment of the clean 
and easily measured Sc and Mn lines of \hd, in which the features are 
broadened by hyperfine structure (HFS) splitting.  HFS results from 
nucleon-electron spin interactions in odd-$Z$ atoms, splitting 
absorption lines into multiple components.  The multi-component 
structure permits the line to grow to a greater strength before 
saturating.  Without accounting properly for HFS, abundances quoted by 
other studies for elements sensitive to HFS can be severely 
over-estimated.  For example, in \hd, including HFS makes a difference 
of 0.35~dex in the derived \eps{Mn} abundance for the \nion{Mn}{i} 
feature at 4030.76~\AA.  For further discussion on the importance of 
including HFS, see \eg, Gratton 1989\nocite{gratton89}, McWilliam et 
al.\ 1995\nocite{mcw+95}, Ryan, Norris, \& Beers 1996\nocite{rnb96}, 
and Prochaska \& McWilliam 2000\nocite{pm00}.  In our analysis of Sc
and Mn, we relied upon the extensive HFS splitting lists produced by 
\citet{kb95}\footnote{The HFS lists are available at 
{\sf http://kurucz.harvard.edu/linelists/gfhyperall/}.} and our 
resulting linelist is given in  Table~\ref{hd.scmn}.

Many of the $n$-capture features of interest also have HFS and/or 
multiple naturally-occurring isotopes with measured wavelength 
differences ($\Delta\lambda$~$\gtrsim$ 0.01~\AA).  Notes regarding 
whether HFS/isotopic splitting was accounted for in a given feature 
are presented in Column 8 of Table~\ref{hd.analysis}.  Additionally, 
nearly all of the strongest $n$-capture transitions occur in the 
complex blue-UV spectral region ($\lambda$~$\lesssim$ 4500~\AA), where 
blended features are the rule, not the exception.  Therefore for more 
than half of the $n$-capture lines we employed full synthetic spectra 
in the analyses.  In most cases a spectral region of 4--8~\AA\ 
surrounding each synthesized line was considered.  As in our prior 
studies, we initially formed the line lists for the synthetic spectra 
from the neutral and first ionized atomic transitions with lower 
excitation potentials $\chi$~$<$ 7~eV found in the \citet{kurucz98} 
atomic and molecular line database\footnote{
Available at http://kurucz.harvard.edu/.}.
In spectral regions with significant molecular hydride and cyanogen 
absorption features, parameters for these lines also were taken from 
the Kurucz database, but for hydrides the lower excitation potential 
cutoff was $\chi$~$\sim$ 1.5~eV.  These lines were occasionally 
supplemented by other transitions identified in the solar spectrum by 
Moore, Minnaert, \& Hourgast (1966\nocite{mmh66}).  In accord with our 
previous work on spectrum syntheses of $n$-capture elements, the 
transition probabilities of a few contaminant lines in some spectral 
intervals were adjusted to provide best matches to solar/stellar 
spectra, employing the \citet{kurucz+84} solar flux atlas, and the 
solar model atmosphere of \cite{hm74}.  On rare occasions we added an 
arbitrarily-assumed \nion{Fe}{i} line to account for weak contaminants 
in the wings of the spectral features of interest.

We convolved the raw synthetic spectra generated from these line lists 
with Gaussian functions whose full-width-at-half-maximum (FWHM) values 
matched the observed spectra. We empirically determined a 
FWHM$_{\rm total}$ of 0.10--0.11~\AA\ at $\lambda$~$\approx$ 4000~\AA, 
which is in accord with the expected value from the convolution of the 
known spectrograph resolution and macroturbulent velocity. Then 
synthetic and observed spectra with varying abundances of the elements 
under scrutiny were iteratively compared until the best average 
abundance was found for each feature.

\begin{figure}
\epsscale{0.9}
\plotone{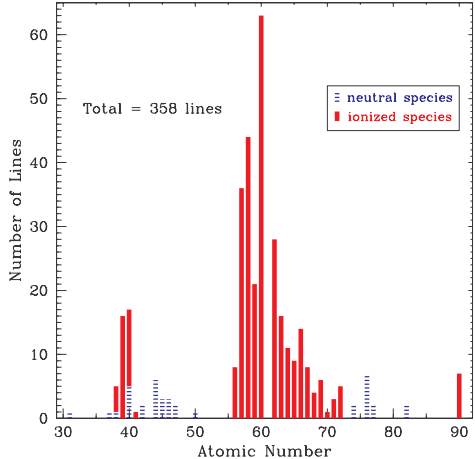}
\caption{
\label{fig.nlines} 
The number of transitions used in determining the abundances of the
$n$-capture elements.  Blue stripes and red solids represent, 
respectively, neutral and first ionized species, as given in the figure 
legend.
}
\end{figure}

More than 350 transitions of 35 species have yielded abundances for 30 
$n$-capture elements and significant upper limits for three others.  
Well-determined abundances involving large numbers of transitions based 
on the best laboratory atomic data require little comment.  In 
Figure~\ref{fig.nlines} we plot a histogram of the number of lines 
employed for each $n$-capture species.  Unsurprisingly, the peak of this 
distribution occurs for the rare earth elements, whose first ions present 
many detectable transitions in the visible and near-UV spectral regions.

\subsection{Comments on Abundance Derivations of Individual Elements}\label{anal-comments}

In this sub-section we concentrate on those species that deserve special 
attention, either to support our abundance claims or to provide some 
cautionary statements.  Only brief notes will be given for most elements. 
However, since the claimed significant departures of HD~221170 
$n$-capture abundances from scaled solar-system values rest mainly on the 
elements Hf, Pb, and Th, they will receive more extended discussion.

{\it CNO group (Carbon: $Z$ = 6; \nion{C}{i} -- Nitrogen: $Z$ = 7; \nion{N}{i} -- Oxygen: $Z$ = 8; \nion{O}{i}):}
Abundances of the CNO group provide clues to the nucleosynthetic history
of \hd.  Additionally, the molecular species formed from these elements
are important contributors to the overall line absorption surrounding
and blending with other features of interest in this study, such as Th.
The C and N abundances here were derived from the CH $A^{2}\Delta-X^{2}\Pi$ 
G-band and the CN $B^{2}\Sigma^{+}-X^{2}\Sigma^{+}$ (0,0) vibrational band.  
We derive a carbon isotope ratio of $^{12}$C/$^{13}$C = 7 $\pm$ 2, in 
agreement with \Yu, who noted their agreement with \citet{spv86}.  Our 
oxygen abundances were derived from full synthetic spectrum computations of 
the $\lambda\lambda$6300, 6363~{\rm \AA}\ [\nion{O}{i}] lines.  The 
construction of the linelists employed are as described by 
\citet{sneden+91b} and \citet{westin+00}.

\begin{figure}
\epsscale{0.9}
\plotone{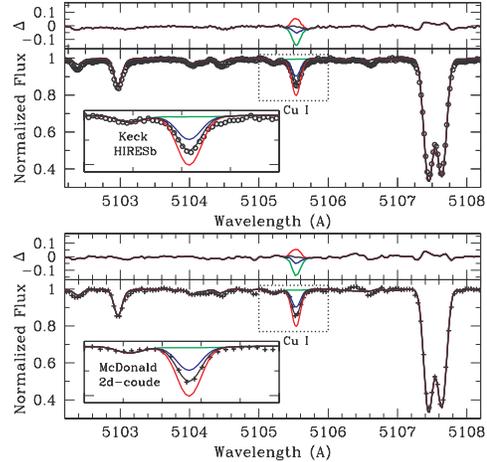}
\caption{
\label{fig.cu} 
Observed and synthetic spectra of \nion{Cu}{i} in \hd.  The 6~\AA\ 
swaths shown here illustrate the typical match between both the 
Keck HIRESb ($\bigcirc$) and McDonald 2d-coud\'e ($+$) spectra, and 
the syntheses obtained from the stellar and smoothing parameters 
described in \S~\ref{anal-teff} and \ref{anal-anal}.  The $\Delta$ 
panels above each of the syntheses illustrate the differences to the 
respective fits.  The inset illustrates a region 1~\AA\ wide 
surrounding the feature at 5105.5~\AA, as denoted by the dotted lines 
in the main panels.  The solid lines represent synthetic spectra with 
variation only in the assumed Cu abundance.  The black line shows the 
best fit between the synthetic and observed spectra.  The red and 
blue lines show the change in the \nion{Cu}{i} feature with changes 
of $\pm$0.2~dex in assumed Cu abundance, respectively, and the green 
lines show the synthetic spectrum without any Cu contribution.  
}
\end{figure}

{\it Copper ($Z$ = 29; \nion{Cu}{i}):} 
The copper abundance for \hd\ was investigated employing the techniques 
and linelists employed in the extensive globular cluster giant star 
study of \citet{jensim+03}.  Unfortunately, the Cu feature at 5782~\AA\ 
could not be utilized here.  The large radial velocity of \hd\ shifts 
the 5782~\AA\ feature into the well-known 5780~\AA\ diffuse interstellar 
band (see \eg, Herbig 1975\nocite{herbig75}).  The small-scale structure 
of this broad band compromised our synthetic spectrum fit to the profile 
of the \nion{Cu}{i} line, so we dropped it from further consideration.
Our Cu abundance therefore is derived exclusively from the spectrum 
synthesis match to the 5105~\AA\ feature, as displayed in 
Figure~\ref{fig.cu}.

{\it Rubidium ($Z$ = 37; \nion{Rb}{i}):} 
Only the resonance line of \nion{Rb}{i}
at 7800.29~\AA\ is strong enough to attempt analysis in metal-poor stars.
We adopted the {\it gf}-value recommended by \citet{fw05}.
Our claimed detection of this very weak feature should be viewed with
caution, as we do through the assignment of a large abundance uncertainty 
value. 

{\it Strontium ($Z$ = 38; \nion{Sr}{i}, \nion{Sr}{ii}):}
The resonance line of \nion{Sr}{i} at 4607.3~\AA\ is detected and 
apparently suffers from little blending in the HD~221170 spectrum.  
Adopting the {\it gf}-value of \citet{mb87}, synthesis of this line 
yields an abundance about 0.4~dex lower than that determined from four 
lines of \nion{Sr}{ii}.  This problem has been noted previously in 
stellar and solar spectra.  The reader is referred to \citet{gs94} for 
further discussion.  Our adopted \nion{Sr}{ii} {\it gf}-values are
those recommended by \citet{fw05}.

{\it Yttrium ($Z$ = 39; \nion{Y}{ii}):} The initial analysis based on EW
measurements yielded an unacceptably-large line-to-line abundance scatter.
Therefore our final values were derived from synthetic spectra, with
full accounting of HFS in the \nion{Y}{ii} lines.
The hyperfine constants were taken from \citet{dinneen+91}; Nilsson,
Johansson, \& Kurucz (1991\nocite{njk91}); \citet{villemoes+92};
\citet{wannstrom+94}; and \citet{persson97}.
The resulting 3--4 components to each line have only small
wavelength differences (always $\lesssim$0.003~\AA), so changes to
the derived Y abundances were less than or comparable to basic
measurement uncertainties in our spectra.
However, better estimation of blending features to the \nion{Y}{ii} 
features brought reasonable internal abundance agreement, 
$\sigma$~=~0.07, among the 16 lines of this study.  All 
{\it gf}-values employed here are adopted from the 
\citet{hannaford+82} study.

{\it Zirconium ($Z$ = 40; \nion{Zr}{i}, \nion{Zr}{ii}):} 
\Yu\
detected \nion{Zr}{i} transitions in HD~221170, the first time this 
species has been identified in very metal-poor stars.  Employing the 
{\it gf}-values from \citet{biemont+81}, we confirm the existence 
of the two lines Yushchenko et al.\ found, add three more lines 
detectable on our spectra, and derive a mean Zr abundance (albeit 
with relatively large line-to-line scatter) that is in excellent 
agreement with the value determined from \nion{Zr}{ii}.

{\it Niobium ($Z$ = 41; \nion{Nb}{ii}):} 
This abundance has been derived from a single detectable line, which 
unfortunately lies in a crowded near-UV spectral region where also the 
stellar flux is relatively low.  We adopted the {\it gf}-value from 
\citet{hannaford+85}. The line has broad HFS, which we accounted for 
empirically by splitting it into a number of substructure components to 
approximately match an emission-line profile from the National Solar 
Observatory  Fourier transform spectrometer (FTS) archives.  A larger 
error bar was assigned to this abundance.

\begin{figure}
\epsscale{0.9}
\plotone{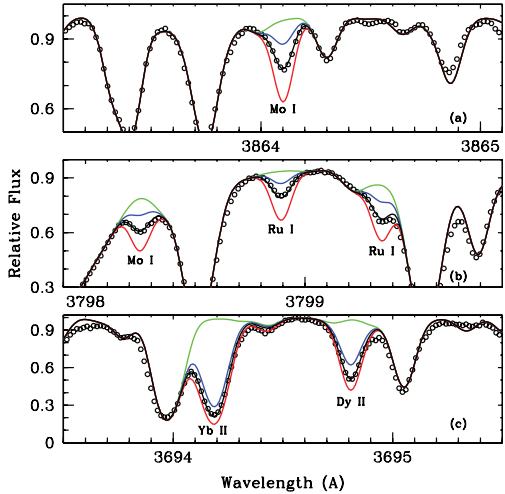}
\caption{
\label{fig.specMo} 
Selected spectra of $n$-capture elements in the near-UV wavelength region. 
In panel (b), the \nion{Mo}{i} line is clearly detected, but lies in the 
middle of a complex blend of \nion{Fe}{i} and \nion{H}{i} features.  The 
abundance from this line is consistent with the cleaner 3864.1~\AA\ 
\nion{Mo}{i} line, but should be interpreted with caution.
}
\end{figure}

{\it Molybdenum ($Z$ = 42; \nion{Mo}{i}):} 
We adopted the {\it gf}-values of \citet{wb88}.  In \S3.2.1 of 
\citet{sneden+03}, there is an extended discussion of the detection and 
analysis of the 3864.10~\AA\ line in CS~22892-052.  The observed 
feature can be attributed to a combination of \nion{Mo}{i}, CH, CN, and 
other atomic transitions, but Sneden \etal\ argued that a reliable Mo 
abundance could be obtained for that star.  The molecular contaminants 
for the 3864~\AA\ line are weaker in HD~221170 than they are in 
CS~22892-052 because its C abundance is smaller, and the atomic 
contaminants also appear to be not strong in HD~221170.  Confirmation 
of the presence of Mo comes from the detection of \nion{Mo}{i} 
3798.25~\AA.  This line lies too close to the \nion{H}{i} Balmer line 
at 3797.90~\AA\ to yield a trustworthy abundance by itself, but our 
best estimate is very consistent with the abundance derived from the 
3864~\AA\ line, bolstering our confidence in the mean Mo abundance.
Both of our \nion{Mo}{i} features are displayed in panels (a) and (b)
of Figure~\ref{fig.specMo}.

{\it Ruthenium ($Z$ = 44; \nion{Ru}{i}):} 
The {\it gf}-values for all six lines were adopted from 
\citet{wickliffe+94}.  Two representative features are displayed in 
panel (b) of Figure~\ref{fig.specMo}.

{\it Rhodium ($Z$ = 45; \nion{Rh}{i}):}  
The lines at 3692.4 and 3700.9~\AA, with transition probabilities from 
\citet{dl85}, yield abundances that are in poor agreement: 
log~$\epsilon$~= $-$0.55 and $-$0.10, respectively.  A relatively clean 
line in the HD~221170 spectrum at 3434.9~\AA\ not done by Duquette \& 
Lawler was included in the \citet{kwiatkowski+82} lab investigation.
The transition probability scales of these two studies are in good 
agreement: for nine lines in common, $<$$\Delta$log~$gf$$>$ = 
$+$0.03~$\pm$~0.04 ($\sigma$~= 0.11) in the sense Duquette \& Lawler 
$minus$ Kwiatkowski \etal\  Adopting the latter's log~$gf$ yields 
log~$\epsilon$~= $-$0.40 for the 3434.9~\AA\ line.

{\it Palladium ($Z$ = 46; \nion{Pd}{i}):}  
The {\it gf}-values were adopted from \citet{biemont+82}.

{\it Silver ($Z$ = 47; \nion{Ag}{ii}):}  
The resonance lines have reliable transition probabilities.
We adopted the values recommended by \citet{fw05}, which are 
in good agreement with the values determined by \citet{cl67}.
See \citet{ra72} for further references and hyperfine/isotopic
substructure discussion.

{\it Tin ($Z$ = 50; \nion{Sn}{i}):}  
We were able to derive an upper limit for the feature at 
3801.02~\AA\ employing the {\it gf}-value recommended by \citet{fw05}.

{\it Barium ($Z$ = 56; \nion{Ba}{ii}):}  
In previous work by our group on selected Ba features we have 
performed blended-line EW analyses which included both hyperfine 
and isotopic subcomponents adopted from \citet{mcw98}, with 
values of \loggf\ normalized to those adopted in our previous work 
(\eg, Ivans et al.\ 1999\nocite{ivans+99}).  As discussed in 
\S~\ref{anal-anal}, however, in \hd\ the $n$-capture material is 
sufficiently enhanced that blending from other $n$-capture lines 
becomes significant in many features (\eg, in the \nion{Ba}{ii} 
4130~\AA\ feature from \nion{Ce}{ii}).  A blended-line EW analysis 
of Ba in \hd\ was therefore deemed untrustworthy and a full spectrum 
synthesis was performed on each feature.  The transition 
probabilities employed here are slightly different than those 
adopted in our previous papers.
A critical compilation of transition probabilities for both \nion{Ba}{i} 
and \nion{Ba}{ii} has been published by Klose, Fuhr, \& Wiese 
(2002\nocite{kfw02}), and
we adopt their recommended values.

We attempted to detect \nion{Ba}{i} 5535.5~\AA, the only strong line of 
this species.  Unfortunately, this line is dominated in the solar 
spectrum by a pair of \nion{Fe}{i} transitions, and the Ba contribution 
to the blend could not be distinguished in our HD~221170 spectrum.

{\it Lanthanum ($Z$ = 57; \nion{La}{ii})}: 
Some features of \nion{La}{ii} have notable HFS and required updated 
calculations to be made of the HFS patterns.  In the Appendix, we 
provide the results of these calculations for 30 of the 36 
\nion{La}{ii} features employed in this study.  Although upper level 
hyperfine constants are not available for the remaining six, the 
profiles of these six lines were inspected on high resolution 
laboratory FTS data from \citet{lawler+01a}.  The broadening induced 
by HFS for these lines is not as large as the broadenings of the other 
30 \nion{La}{ii} lines, and are not sufficiently resolved in the FTS 
data to derive a precise measure of the HFS constants.  The effect of 
the (weak) HFS splitting for these lines was, however, included 
(approximately) in the analysis of the stellar spectra.  The complete 
HFS patterns from the best available HFS constants are included for 
the other 30 \nion{La}{ii} lines in the Appendix.  

{\it Cerium ($Z$ = 58; \nion{Ce}{ii}):}  
Abundances derived for all 44 lines relied upon the {\it gf}-values
of \citet{palmeri+00}

{\it Praseodymium ($Z$ = 59; \nion{Pr}{ii}):}  
All 21 features were analysed employing the {\it gf}-values of
\citet{ivarsson+01}.

{\it Neodymium ($Z$ = 60; \nion{Nd}{ii}):}  
We performed 63 abundance measurements employing the 
extensive linelist of \citet{denhartog+03}.

{\it Samarium ($Z$ = 62; \nion{Sm}{ii}):}  
Accurate experimental log~{\it gf}-values 
have been newly determined for over 900 \nion{Sm}{ii} lines by 
\citet{lawler+05}.  The reader is referred to that paper for new 
determinations of the Sm abundances in the Sun and the $r$-process-rich 
stars CS~22892-052, HD~115444, \bd.

{\it Europium ($Z$ = 63; \nion{Eu}{ii}):} 
We employed the {\it gf}-values derived by \citet{lawler+01c}.
Updated energy levels for \nion{Eu}{ii} and complete HFS and 
isotopic patterns are included in Appendix for 24 \nion{Eu}{ii} 
lines.  The important low-lying even- and odd-parity Eu II 
levels were measured to FTS (interferometric) accuracy of $\pm$ 
0.003~\invcm.

For this $r$-process-dominated element, there are two naturally 
occurring isotopes whose abundance fractions are 
fr(\iso{151}{Eu})~$\equiv$ \iso{151}{Eu}/(\iso{151}{Eu}+\iso{153}{Eu})
= 0.478 and fr(\iso{153}{Eu})~= 0.522 in solar-system material (e.g., 
see the review of Lodders 2003\nocite{lodders03}, and references 
therein).  \citet{sneden+02} and \citet{aoki+03} have demonstrated 
that the $r$-process-rich stars HD~115444, \bd, \cs, and CS~31082-001 
also have approximately equal fractions of the two Eu isotopes.
For \hd, we computed synthetic spectra of the 3819.7, 4129.7, and 
4205.0~\AA\ with varying \iso{151}{Eu} and \iso{153}{Eu} fractional 
abundances, and similarly find fr(\iso{151}{Eu})~$\approx$ 
fr(\iso{153}{Eu})~$\approx$ 0.5.  However, since these most useful Eu 
isotopic abundance indicators are very strong in HD~221170 and are 
surrounded by complex atomic and molecular contaminants, this isotopic 
estimate is not tightly constrained, and the uncertainty in each 
abundance fraction is roughly $\pm$0.1.

{\it Gadolinium ($Z$ = 64; \nion{Gd}{ii}):} For these lines, we 
employed the {\it gf}-values of \citet{bergstrom+88}, supplementing 
these with some from the \citet{kurucz98} database consistently 
normalized with the Bergstr\"om et al.\ values.

{\it Terbium ($Z$ = 65; \nion{Tb}{ii}):} The {\it gf}-values
derived by \citet{lawler+01b} were employed for all features.

{\it Dysprosium ($Z$ = 65; \nion{Dy}{ii}):} We 
adopted the {\it gf}-values from \citet{wickliffe+00}.  A representative
feature is displayed in panel (c) of Figure~\ref{fig.specMo}.

{\it Holmium ($Z$ = 67; \nion{Ho}{ii}):} The abundances were 
derived employing the {\it gf}-values from \citet{lawler+04}.

{\it Erbium ($Z$ = 68; \nion{Er}{ii}):} We adopted the
{\it gf}-values of \citet{xu+03}, preferring to limit our transition
list to only the relatively few reported in that study rather than 
the larger numbers of \nion{Er}{ii} lines included in older laboratory
studies.  Of the many \nion{Er}{ii} lines, these are the only ones 
with sufficiently up-to-date parameters to be included in The Database 
of Rare Earths at Mons University (D.R.E.A.M.).\footnote{The database 
is available at {\sf http://w3.umh.ac.be/~astro/dream.shtml}.}  In 
addition, Er has four major naturally-occurring isotopes whose HFS
splitting has not been taken into account in any stellar abundance
analyses to date.  Thus, Er abundances are likely to have been 
overstated with the reported associated uncertainties, understated.

{\it Thulium ($Z$ = 69; \nion{Tm}{ii}):} Our abundance was
derived employing the linelist of \citet{wl85}.

{\it Ytterbium ($Z$ = 70; \nion{Yb}{ii}):} The {\it gf}-value 
was adopted from \citet{biemont+02}.  The feature is displayed in
Figure~\ref{fig.specMo}.

{\it Lutetium ($Z$ = 71; \nion{Lu}{ii}):} The upper limit
was derived from three lines where {\it gf}-values from \citet{fedchak+00}
were employed.

{\it Hafnium ($Z$ = 72; \nion{Hf}{ii}):} 
\Yu\ analyzed the
3918.09 and 4093.16~\AA\ lines, and derived abundances that differed by
0.8~dex in log~$\epsilon$, or 0.5~dex in [Hf/H].  In their \S3 they 
discussed this disagreement, arguing for its reality from significant 
EW differences: they measured EW$_{3918}$~= 45~m\AA\ and EW$_{4093}$~= 
15~m\AA.  They chose to use the mean abundance derived from these two 
lines for their final Hf value.

\begin{figure}
\epsscale{0.9}
\plotone{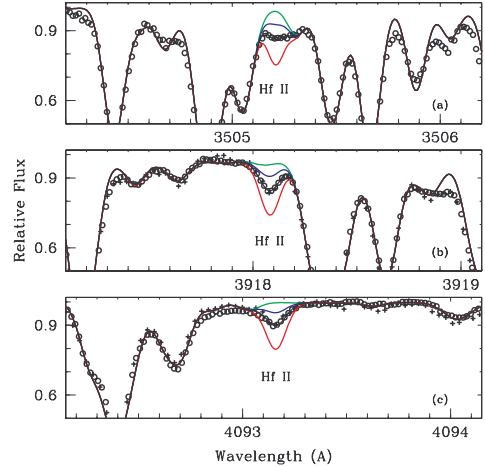}
\caption{
\label{fig.specHf} 
Observed and synthetic spectra of \nion{Hf}{ii} lines in \hd.
Open circles and crosses represent the observed Keck HIRESb 
and McDonald 2d-coud\'e spectra, respectively. The solid
lines represent synthetic spectra with variation only in
the assumed Hf abundance.  The black line shows the best 
fit between the synthetic and observed spectra with
the abundance for each line given in Column 6 of 
Table~\ref{hd.analysis}.  
The red and blue lines show the change in the \nion{Hf}{ii} 
lines with changes of $\pm$0.4~dex in assumed Hf abundance, 
respectively, and the green line shows the synthetic 
spectrum without any Hf contribution.
}
\end{figure}

We have compared synthetic and observed spectra for these two lines and
three other \nion{Hf}{ii} lines detected in our spectra, and in 
Figure~\ref{fig.specHf} we display the data for the three most 
trustworthy features.  For the 3918 and 4093~\AA\ lines, panels (b) and 
(c) respectively, we have Keck and McDonald observed spectra of 
comparable quality, so both are plotted.  The S/N of the McDonald 
spectrum is too low at 3505~\AA\ to be useful in the analysis.  Panel 
(c) of this figure can be compared to Figure~1 of Yushchenko \etal\  
The \nion{Hf}{ii} lines in Figure~\ref{fig.specHf} are all about the 
same strength, as they should be, given their excitation potentials and 
oscillator strengths listed in Columns 2--3 of Table~\ref{hd.analysis}.  
The {\it gf}-values were adopted from \citet{andersen+76}.  The EWs of 
the 3918 and 4093~\AA\ lines both are $\sim$15~m\AA.  Our derived 
abundance for the 4093~\AA\ transition is in excellent agreement with 
the Yushchenko \etal\ value.  Some problem probably exists in their 
spectrum of the 3918~\AA\ feature, and the small line-to-line scatter 
among the five lines employed in the present study lends some confidence 
in our derived mean Hf abundance, which is $\sim$0.3--0.4~dex smaller 
than that derived by Yushchenko \etal\

{\it Tungsten ($Z$ = 74; \nion{W}{i}):}  
\Yu\ employed lines at 4008.7 and 4074.4~\AA\ in their analysis.  We 
also computed synthetic spectra of these lines, which should be the 
strongest of this element in the HD~221170 spectrum.  We conclude that 
measurable \nion{W}{i} absorption is probably present but both lines 
are so contaminated with other spectral features that it is difficult 
to derive a meaningful abundance.  Here we will describe the estimation 
of an upper limit.

The 4008.75~\AA\ \nion{W}{i} line is heavily blended at least by 
\nion{Pr}{ii} 4008.69~\AA, \nion{Ce}{ii} 4008.67, 4008.73~\AA, and 
\nion{Nd}{ii} 4008.75~\AA\ (this line is of relatively high excitation 
potential, and thus unlikely to contribute substantially).  Inspection 
of the feature profile suggests that perhaps three transitions make up 
the total absorption.  As might be expected, the overall spectral 
feature strength increases and decreases in rough proportion to the 
overall $n$-capture element content.  Our syntheses suggest that the 
contaminants account for about 3/4 of the total feature.  The 
\nion{W}{i} line is probably contributing to the absorption, and we 
formally derive log~$\epsilon$(W)~$\sim$~$-$0.6, but the uncertainty 
in the W abundance derived from spectra with the resolution and S/N of 
our spectra must be very large ($\sim\pm$0.4~dex).  The two 
\nion{Ce}{ii} lines have recently-determined transition probabilities 
\citep{palmeri+00}.  The \nion{Pr}{ii} line was not included in the 
\citet{ivarsson+01} lab study, and so we experimented with syntheses 
that did not include it.  The observed feature is clearly not matched 
in this case, with an obvious gap at the \nion{Pr}{ii} line wavelength.

The 4074.36~\AA\ line is approximately a factor of two weaker than the
4008~\AA\ line. It too suffers blending, from the CH 
$B^2\Sigma^-$--$X^2\Pi$ (1--1) P$_1$ J~=~4.5 $\lambda$4074.34~\AA\ line 
(e.g., Bernath \etal\ 1991\nocite{bernath+01}).  Repeated trial 
syntheses, assuming the C abundance derived from the CH 
$A^2\Delta$--$X^2\Pi$ G-Band, suggest that the CH feature dominates 
here, and so log~$\epsilon$(W)~$\lesssim$~$-$1.0.  Complete neglect of 
the CH contaminant would lead to log~$\epsilon$(W)~$\approx$~$-$0.6.
We conclude that \nion{W}{i} lines may have been detected in the
spectrum of HD~221170, but probably cannot at this time be employed
as reliable W abundance indicators.  A conservative upper limit is 
log~$\epsilon$(W)~$\lesssim$~$-$0.6, based on the {\it gf}-values of 
\citet{denhartog+87}.

{\it Osmium ($Z$ = 76; \nion{Os}{i}):}  
\citet{ivarsson+03} transition
probabilities were employed except for the 4420.5~\AA\ line, for which
the log~{\it gf}-value was determined by \citet{kwiatkowski+84}.  We 
included this line because Ivarsson et al.\ also used it in their 
determination of the Os abundance of CS~31082-001.

{\it Iridium ($Z$ = 77; \nion{Ir}{i}):}  
Our analysis relied upon the {\it gf}-values of \citet{ivarsson+03}.

{\it Lead ($Z$ = 82; \nion{Pb}{i}):}  
\Yu\ derived their lead abundance from the 4057.8~\AA\ line.  We have 
also used this transition along with that at 3683.5~\AA,  with both 
{\it gf}-values adopted from \citet{biemont+00}.  The resulting 
abundances are in fair agreement, considering the weakness of the two 
lines and the blending problems of each (see Figure~2 of Yushchenko 
et al). 

{\it Thorium (Z~=~90, \nion{Th}{ii}):}  
\citet{yu+02} were the first to report on the abundance of Th in this 
star.  In their expanded follow-up, \Yu\ employed seven lines.  Five 
of these lines (4019.1, 4086.5, 4094.7, 4179.7, and 5989.0~\AA) were 
included in the \citet{nilsson+02} lab study of \nion{Th}{ii}, but 
the 4003.3 and 4178.1~\AA\ lines were not.  Therefore Yushchenko 
\etal\ quoted mean Th abundances both from the five lines with 
Nilsson \etal\ log~{\it gf}-values and from the whole set of seven 
lines.  Of their five lines with Nilsson \etal\ data, three must be 
viewed with caution.  The 4094.75~\AA\ line is present in the \hd\
spectrum but it is blended with \nion{Er}{ii} 4094.64~\AA, CH 
4094.70~\AA, and \nion{Fe}{ii} 4094.73~\AA.  The 4179.71~\AA\ line is 
intrinsically very weak, and it lies between (and is mostly masked 
by) \nion{Nd}{ii} 4179.58~\AA\ and \nion{Zr}{ii} 4179.81~\AA.  It is 
useful only to estimate an upper limit of the Th abundance.  The 
5989.05~\AA\ line occurs in a spectral region of significant telluric
H$_2$O contamination, which must be removed from the \hd\ spectrum
via division by the spectrum of a rapidly-rotating hot star.

\begin{figure}
\epsscale{0.9}
\plotone{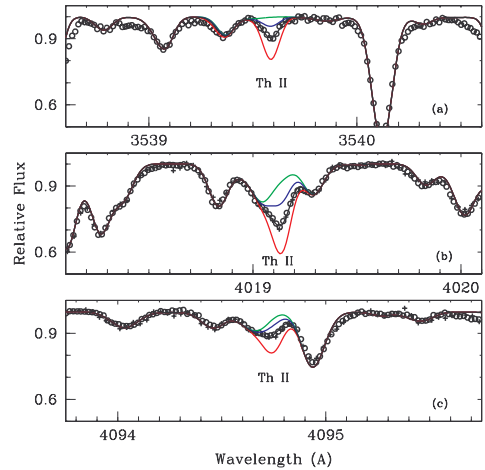}
\caption{
\label{fig.specTh} 
Observed and synthetic spectra of three \nion{Th}{ii} lines in HD~221170.
All lines and symbols are as in Figure~\ref{fig.specHf}.
}
\end{figure}

In the present study we have chosen to consider only \nion{Th}{ii} 
lines with transition probabilities determined by Nilsson \etal\
Our search for lines in the HD~221170 spectrum yielded seven total, 
four in common with Yushchenko \etal\  In Figure~\ref{fig.specTh} we 
show synthetic and observed spectra for three lines.  Panel (a) 
displays an apparently clean feature at 3539.6~\AA.  We have 
discovered no plausible identification other than \nion{Th}{ii} at 
this wavelength.  In panel (b) we show the 4019.1~\AA\ line, which 
in previous studies often has served as the sole Th abundance 
indicator.  This line appears to be relatively clean in extremely 
$r$-process-rich stars such as CS~31082-001 (see Figure~9 of Hill 
\etal\ 2002\nocite{hill+02}).  However, in higher metallicity, less 
extreme $r$-process stars, there are well-known blending issues that 
increase the Th abundance uncertainty from this complex blend.  We 
have modeled the total feature as well as possible, trying to 
account for $^{\rm 13}$CH, \nion{Ce}{ii}, \nion{Fe}{i}, and 
\nion{Co}{i} contaminants, but the resulting abundance here should 
be viewed with caution.  Finally, in panel (c) we show the 
4094.7~\AA\ line.  The observed feature on our spectra clearly is a 
blended one, with a FWHM that is larger than any neighbouring single 
line.  This illustrates our contention that it is substantially more 
blended than is the 4019.1~\AA\ line.  The Th abundance derived from 
this feature is very sensitive to the assumed strengths of the 
contaminating transitions listed in the preceding paragraph.

The final Th abundance is taken here as a straight mean of the results for
all seven \nion{Th}{ii} transitions.
There are various blending issues associated with most of these lines,
and only the 3539.6 and 4019.1~\AA\ lines are strong enough to dominate
the surrounding contaminants.
But the synthetic/observed spectrum matches yield nearly the same abundance
from all lines, lending confidence to the derived mean.

\section{ABUNDANCE RESULTS\label{anal-abtot}}

Table~\ref{hd.analysis} includes the derived abundance for each 
feature, in \eps{X} (Column 6) and in bracket notation relative to the 
scaled solar value (Column 7), where the abundance for individual 
lines is displayed.  For iron, the bracket notation values are [Fe/H],
and for all other elements X the values are [X/Fe].  In 
Table~\ref{hd.abtot}, we present a summary of the abundances derived 
from both EW measurements and spectrum syntheses. The first column 
denotes whether the abundance is from a neutral or ionized species, 
or if it represents the mean.  The mean abundance was calculated by 
treating all lines of a given species with equal weight.  The value 
of $\sigma_{lines}$ denotes the $\sigma$ derived from the line-to-line 
scatter.  The value of $\sigma_{adopted}$ includes an allowance for 
any uncertainty in the spectroscopically derived parameters.  
Abundances derived from a single feature have been assigned a minimum 
$\sigma$ = 0.2~dex, and for species represented by only two lines, we 
have assigned a minimum $\sigma$ = 0.1~dex.

For species represented by less than about five or fewer lines, 
uncertainties associated with individual transitions (blending, 
continuum placement, transition probabilities, hyperfine/isotopic 
substructure) are the limiting factors in the mean abundances. 

As emphasized by the referee and noted by \citet{en03}, continuum 
placement can be an unaccounted-for source of error in some abundance 
analyses.  Erspamer \& North find that continuum placement errors as 
small as one percent can have a large affect on the abundance 
results of some stars obtained from even high S/N, high resolution 
data such as those gathered with the ELODIE \'echelle spectrograph. 
Table 2 and Figure 3 of their study of A -- F stars show that 
abundance differences of up to a few tenths of a dex can occur in 
stars possessing $V$$\sin{i}$ of 150 \kmsec.  Fortunately,  the 
$V$$\sin{i}$ of our K star HD221170 is lower than the lowest 
$V$$\sin{i}$ value presented by Erspamer \& North, for which they 
determined that abundance differences resulting from systematic 
uncertainties in continuum placement of $\pm$~1\% were all less 
than 0.1~dex.  For species with more transitions, these concerns 
are of less importance; systematic effects (scale uncertainties in 
transition probabilities, model atmosphere parameter choices, etc.) 
begin to dominate.

\begin{figure}
\epsscale{0.9}
\plotone{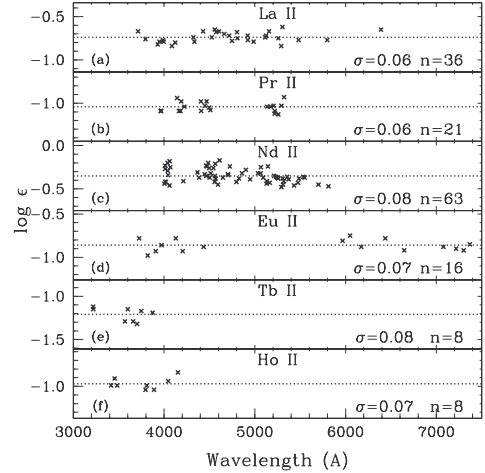}
\caption{
\label{fig.lambda} 
Abundances from individual lines of six rare-earth element species,
plotted against line wavelengths.  The mean abundances are indicated 
by dotted horizontal lines, with the scatter about the mean ($\sigma$) 
and the number of lines employed ($n$) stated in each panel.
}
\end{figure}

To illustrate the line-to-line scatter for some of the $n$-capture 
elements in the HD~221170 spectrum, in Figure~\ref{fig.lambda} we plot 
abundances from individual transitions as functions of wavelength, for 
six species with at least eight transitions apiece.  The chosen 
species are: \nion{Nd}{ii}, with the largest number of lines in our 
analysis; \nion{La}{ii}, \nion{Pr}{ii}, \nion{Tb}{ii}, and 
\nion{Ho}{ii}, rare earths whose abundances must be derived from 
synthetic spectra that take account of HFS: and \nion{Eu}{ii}, which 
has significant hyperfine and isotopic splitting as well.  The number 
of lines and the standard deviations ($\sigma$) are noted in each panel 
of the figure.  Several of these elements have useful lines spanning a 
large wavelength range ($\Delta\lambda$~$\gtrsim$ 1000~\AA).  However, 
Tb and Ho (panels (e) and (f), respectively) have detectably strong 
transitions only at shorter wavelengths ($\lambda$~$\lesssim$~ 
4000~\AA).  It is clear from this figure that the derived abundances 
have no significant dependence on wavelength.  The abundances also 
show no obvious variations correlated with line excitation potential 
or log~$gf$, but the baselines in those two quantities are too small to 
warrant further comment.

\section{COMPARISON OF ABUNDANCE RESULTS WITH OTHER STUDIES\label{anal-comp}}

In this section, we compare our derived abundances for \hd\
with those obtained in other studies, as well as with the
expectations from scaled-solar $r$-process pattern 
predictions, and with other $r$-process-rich stars.

\subsection{Comparison to Other Studies of HD~221170}

To date, the most comprehensive abundance studies of HD~221170 have been 
those of \citet{yu+02,yu+05}, and \citet{gopka+04}.  Other abundance
analyses have been made in the context of other programs, with usually only 
a limited set of elemental abundances investigated in a given study.  
\citet{burris+00} analysed \hd\ using data of $R$ $\simeq$ 20,000 and a 
wavelength coverage of 4070~\AA\ $<$ $\lambda$ $<$ 4710~\AA, supplemented 
with data surrounding the $\lambda$ 6141~\AA\ \nion{Ba}{ii} feature.  For 
the seven elements of $Z >$ 39 in common, we find 
$<$$\Delta$[X/Fe]$>_{\rm Burris}$ = $-$0.14 ($\sigma$ = 0.13), in the 
sense of this study $minus$ \citet{burris+00}.  \citet{barklem+05} 
included HD~221170 as a comparison object (S/N $\sim$50) in their survey 
of $r$-process enhanced stars taken with $R$ $\sim$ 20,000 and wavelength 
coverage of 3760\AA $<$ $\lambda$ $<$ 4980\AA.  Comparing the same 
elemental range, for seven elements in common, we find 
$<$$\Delta$[X/Fe]$>_{\rm HERES}$ = $-$0.10 ($\sigma$ = 0.03), in the 
sense of this study $minus$ \citet{barklem+05}.  \citet{fulbright00} 
also observed this star.  While we obtain excellent agreement with 
the EWs in common with the Fulbright study 
(recall Figure~\ref{fig.ew}), for the three elements of $Z >$ 39 in 
common, we find $<$$\Delta$[X/Fe]$>$ = 
$+$0.25 ($\sigma$ = 0.19) where the abundance differences are driven 
by the large \vmicro\ (2.75~\kmsec) adopted by Fulbright.  And, in the 
independent analysis by \citet{jensim+04} based on 2d-coud\'e data 
alone, a value of \eps{Eu/La} = $-$0.13 was derived, the same value as 
derived in this study.

For the 35 elements in common with the \citet{yu+05} study, 
$<$$\Delta$[X/Fe]$>_{\rm Yuschenko}$ = $+$0.04 ($\sigma$ = 0.19), in the 
sense of this study $minus$ \citet{yu+05}.  The mean difference between 
the two studies in a direct comparison of [X/Fe] values is statistically 
insignificant.  This difference is fortuitously small, but it is also 
meaningless: the mean abundance difference encodes little or no 
information across 35 elements.  The analysis of the derived abundances 
in HD~221170 involves careful comparisons against predicted abundance 
patterns.  Patterns are examined in the deviations from the mean 
($\sigma$).  Deviations from the mean are of tremendous importance 
because those individual elements can be the basis from which critical 
inferences are made. Before confronting nucleosynthetic predictions, 
one requires a {\em reliable} set of derived abundances.  

\begin{figure}
\epsscale{0.9}
\plotone{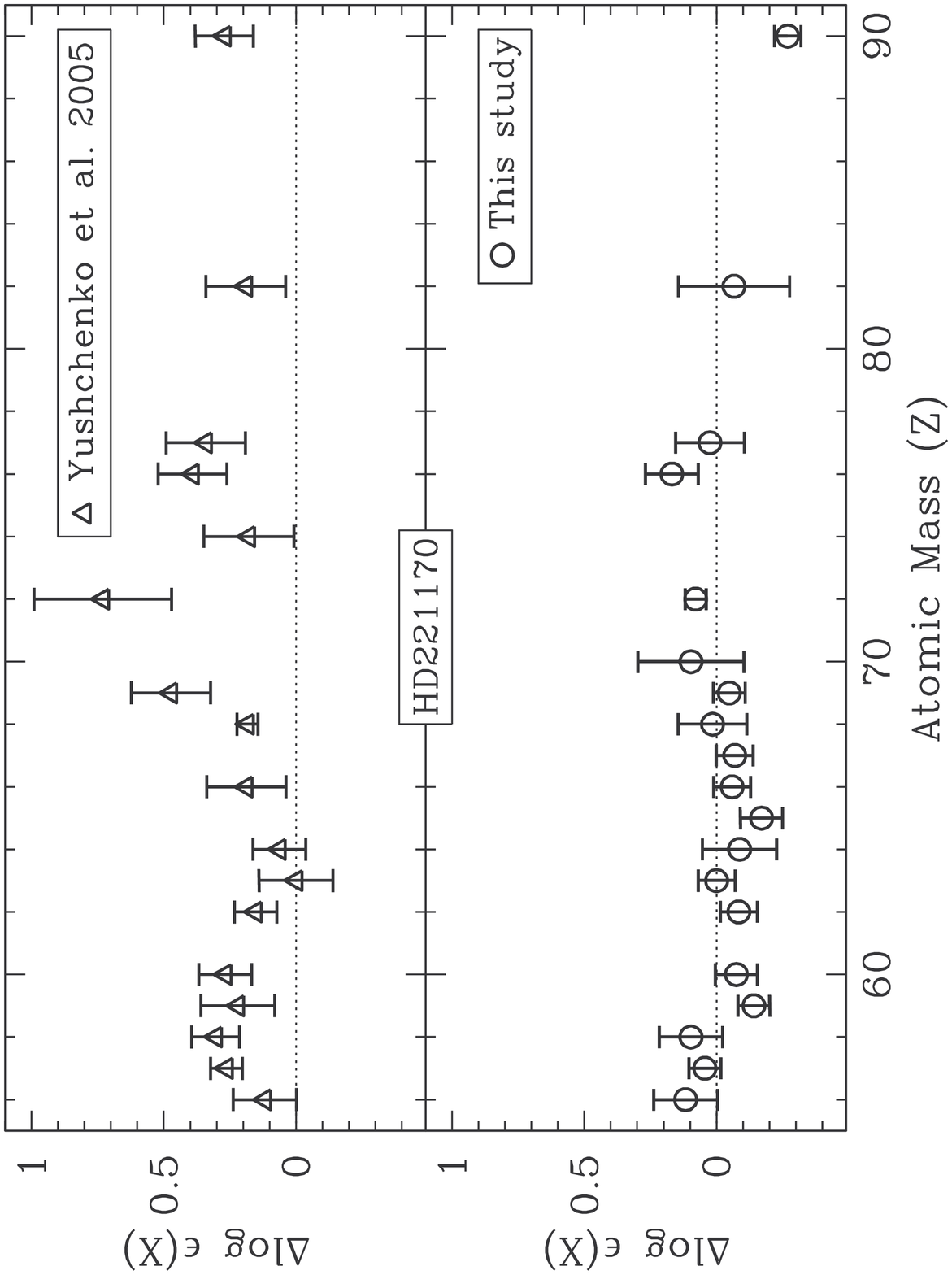}
\caption{
\label{fig.riiiyu} 
\hd\ abundance comparisons of the heavy $n$-capture elements to the 
predicted scaled solar $r$-process pattern of \citet{jensim+04}.  In 
the top panel, we display the difference in the abundances reported 
by \Yu, and in the bottom panel, the differences derived in this 
study.  The derived abundances have been normalized to the value 
derived for \eps{Eu} in the corresponding study.  The error bars 
displayed are those of the 1-$\sigma_{adopted}$ values.
}
\end{figure}

In Figure~\ref{fig.riiiyu}, we display the differences of the heavy 
$n$-capture abundances derived in this and the \Yu\ studies with 
respect to the scaled solar $r$-process predictions of 
\citet{jensim+04}.  

As in our previous papers on $r$-process-rich stars, we have 
normalized the derived abundances to the value derived for \eps{Eu} 
in the corresponding study.  By most accounts, $\sim$95\% of the Eu 
in the Sun was produced by the $r$-process: 93\% \citep{kbw89}; 94\% 
\citep{arlandini+99, travaglio+04}; 97\% (Anders \& Grevesse 
1989\nocite{ag89}, Burris et al 2000\nocite{burris+00}; as also 
reported in Simmerer et al 2004\nocite{jensim+04}). 

As \Yu\ note, their derived $n$-capture abundance pattern for 
\hd\ normalized to Er does not match well the scaled solar 
$r$-process abundance distribution, particularly for elements 
with $Z >$ 68.  In addition to the difficulties of analyzing
Er (see \S~3.4), there are other potential pitfalls in 
normalizing to Er.  Only $\sim$85\% of the solar Er is produced 
via the $r$-process (see references given in \S~1).  
The $r$-process contribution to Er less resembles Eu, Ir, Pt, 
and Au than it does Bi, Gd, Tm, and Dy.  In studies of 
metal-poor stars with $s$$+$$r$ enhancements (see \eg\ Aoki et 
al 2002\nocite{aoki+02}; Johnson \& Bolte 2004\nocite{jb04}),
[Er/Eu] can be $\sim$1 {\em regardless} of whether or 
not the stars possess strong $r$-process enhancements.

In Figure~\ref{fig.riiiyu}, the derived abundances of both 
studies have both been normalized to the value derived for 
\eps{Eu} in the corresponding study.  
In contrast to the results of \citet{yu+05}, in this atomic 
mass range, we find 
good agreement between the abundances derived here and the 
predicted scaled solar $r$-process values.  Our derived Th 
abundance is discussed further in \S~7.  For some other 
elements, there is a disagreement between the derived solar 
photospheric and chondritic/meteoritic abundances.  For instance,  
\citet{lodders03} notes discrepancies of $-$0.07, $+$0.11 and 
$+$0.08~dex for Pr ($Z$ = 59), Hf ($Z$ = 72) and Os ($Z$ = 76), 
respectively.  Incorporating additional uncertainties such as these in 
the {\em solar} abundance scale is sufficient to push our observed and 
predicted abundances into good agreement. 

Many of the abundance differences between our results and those of 
\Yu\ appear to arise from differences in the quality of the spectra 
employed in the two studies.  Some of the most obvious differences can 
be detected by eye.  For instance, a comparison can be made between 
the spectra displayed in the panels of our Figure~\ref{fig.spectra} 
against the spectra shown in Figures 2 of Yushchenko et al.\ 
(2002\nocite{yu+02}) and Figures 1--4 of \Yu.  As already noted in 
\S~2, the appearance of the spectra employed by Yushchenko et 
al.\ is quite different from either the Keck HIRES or McDonald 
2d-coud\'e data employed here.  Furthermore, as shown in our 
Figure~\ref{fig.ew}, there are significant differences in the EWs 
measured by \Yu, not only against those of this study measured in 
Keck HIRES and McDonald 2d-coud\'e data, but also those of yet 
another independent set of measurements employing Lick Hamilton data 
\citep{fulbright00}. In addition, there are differences in the atomic
parameters employed in the two studies (e.g., for elements such as Sm, 
Nd, Eu, and Th).  The thorium abundance differences, in particular, 
seem to relate not only to differences in the adopted 
log~{\it gf}-values but we note that the differences are largest for 
``redder'' wavelengths.  We suggest that the thorium abundance 
differences may be more related to line strengths and data quality 
than to anything else.  Furthermore, in many cases, the lines employed 
in the \Yu\ study yielded anomalously large abundances due to 
un-accounted-for blends, including but not limited to those arising 
from HFS splitting.  Thus, we are unable to replicate many of the 
$n$-capture abundances deduced by \Yu\ for \hd.

\subsection{Comparison to Scaled Solar $r$-process Predictions for $Z \ge$ 56\label{comp-ss}}

\begin{figure}
\epsscale{0.85}

\plotone{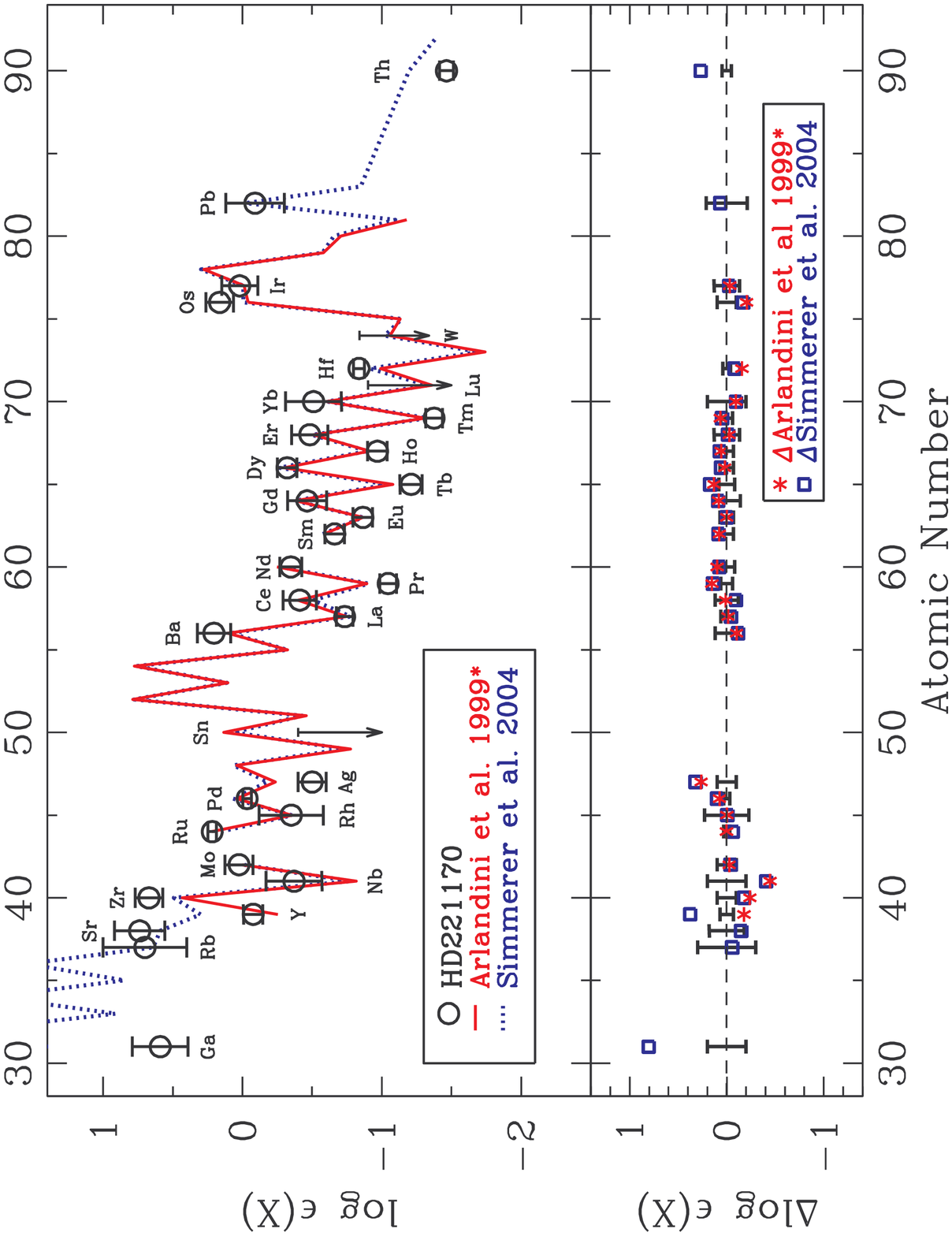}
\caption{
\label{fig.abn} 
Comparison of the \eps{X} abundances for $Z >$ 30 in \hd\ with the 
scaled-solar $r$-process predictions from Simmerer et al.\ 
(2004; solid red line) and those based on Arlandini* et al.\ (largely 
1999; with modifications as described in \S~\ref{comp-ss}).  Both 
sets of predictions have been normalized to the value derived for 
\eps{Eu} in \hd.  In the top panel, the upper limits and open circles 
with error bars denote the stellar abundances.  The bottom panel 
displays the difference defined as $\Delta$\eps{X} $\equiv$ 
\eps{X}$_{\rm pred}$ $-$ \eps{X}$_{\rm obs}$ where the error bars are 
those we adopted for the abundances derived for each element.  Upper 
limits are not displayed in the bottom panel. 
}
\end{figure}

In Figure~\ref{fig.abn} we display the abundances we derived for 
\hd\ in the context of predictions of the scaled solar $r$-process 
abundances by \citet{jensim+04} and Arlandini et al.\ 
(1999\nocite{arlandini+99}; with most values taken from Table~10 of 
Simmerer et al.)  In the study of \bd\ by \citet{cowan+02}, it was 
found that the \citet{arlandini+99} distribution seemed to fit the 
abundances of the $r$-process-rich star better than the predictions 
of \citet{burris+00}. \citet{jensim+04} slightly revised the 
\citet{burris+00} values, and included the updated value of La from 
\citet{obrien+03}, which they also incorporated in their updated 
\citet{arlandini+99} La value.  Furthermore, we take into account 
here the $r$-only isotopic contributions of $^{124}$Sn, $^{130}$Te, 
$^{136}$Xe, and $^{150}$Nd to the total scaled solar $1 - s = r$-process 
abundance predictions to the Arlandini et al.\ values presented in
Simmerer et al.  The resulting predictions are labelled Arlandini* and
are those employed for the remainder of this paper.  

For 13 elements with 56 $\le Z \le$ 69 in \bd, the differences between 
the observed abundances and the scaled solar predictions normalized to 
the Eu abundance value derived for \bd\ ($\Delta$\eps{X} $\equiv$ 
\eps{X}$_{\rm pred}$ $-$ \eps{X}$_{\rm obs}$) are 
$<$$\Delta$(log~$\epsilon$)$>$$_{\rm Arlandini*}$ = 
0.031 $\pm$ 0.019 ($\sigma$ = 0.068) versus 
$<$$\Delta$(log~$\epsilon$)$>$$_{\rm Simmerer}$ = 0.038 $\pm$ 0.024 
($\sigma$ = 0.081).  For the same elemental range, we obtain the 
following differences in the abundances of \hd\ and the predictions of 
the scaled solar $r$-process pattern, normalized to our Eu abundance:
$<$$\Delta$(log~$\epsilon$)$>$$_{\rm Arlandini*}$ = $-$0.042 $\pm$ 
0.019 ($\sigma$ = 0.070) and 
$<$$\Delta$(log~$\epsilon$)$>$$_{\rm Simmerer}$ = $-$0.035 $\pm$ 0.024 
($\sigma$ = 0.086).  Thus, in the case of \hd, the scatter is slightly 
smaller with respect to the Arlandini* et al.\ scaled-solar $r$-process 
abundances, with the offset slightly smaller with respect to the 
Simmerer et al.\ predictions.  The differences are small, but not a
function of metallicity: within the uncertainty, both stars share the
same metallicity ([Fe/H] of \bd\ is $-$2.08 with $\sigma$ = 0.08 and 
that of \hd\ is $-$2.18 with $\sigma$ = 0.12). 

\section{REVISITING THE ISSUE OF MULTIPLE $r$-PROCESS SITES}

\begin{figure}
\epsscale{0.85}

\plotone{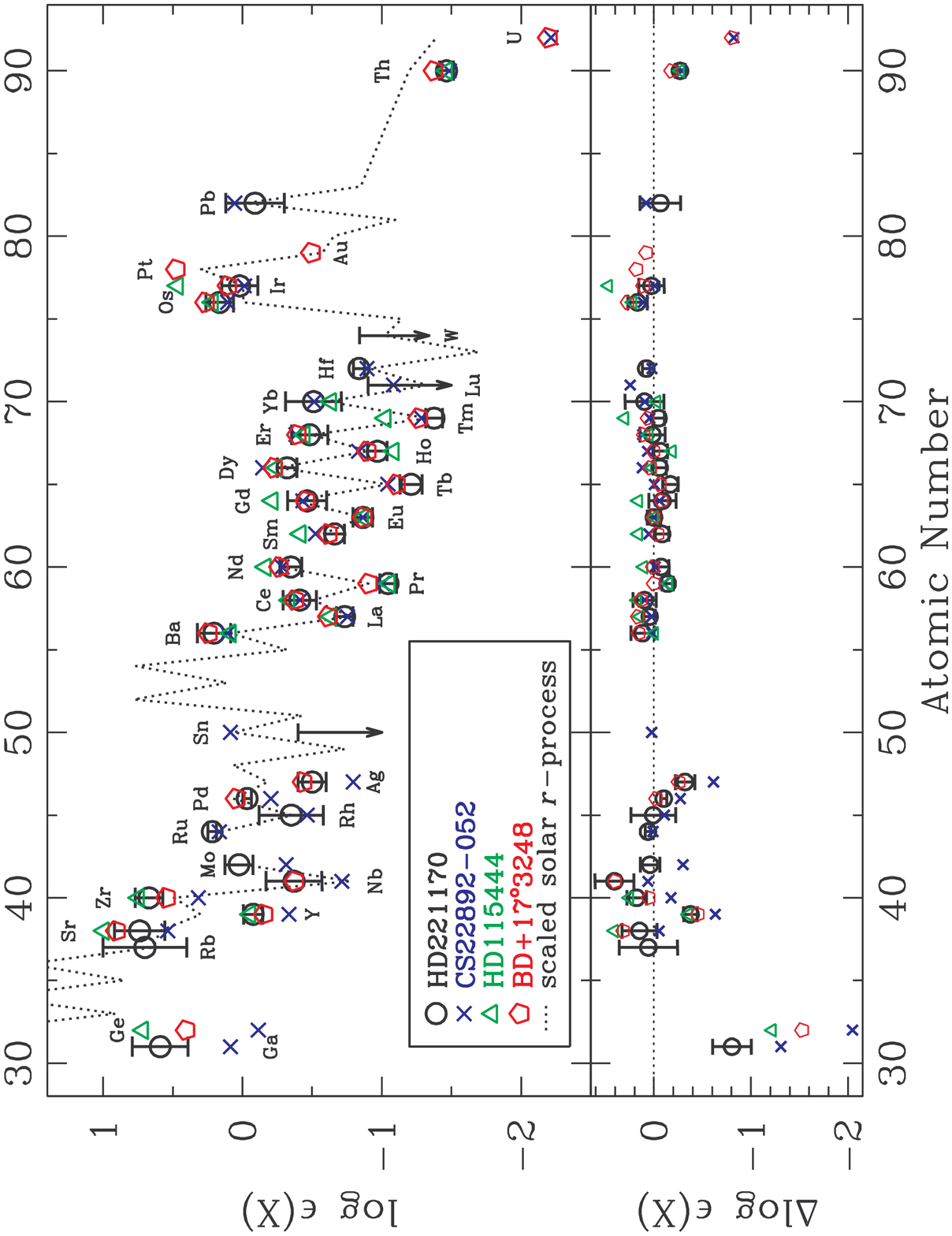}
\caption{
\label{fig.rmultistar} 
Comparison of the \eps{X} abundances of $Z >$ 30 in \hd\ with those of 
CS~22892-052 (blue $\times$; Sneden et al.\ 2003), HD~115444 (green 
$\bigtriangleup$; Westin et al.\ 2000), \bd\ (red pentagon; Cowan et 
al.\ 2002), and the predicted scaled solar $r$-process pattern 
(dotted line; Simmerer et al.\ 2004).  Both the derived and predicted 
abundances have been normalized to the value derived for \eps{Eu} in 
HD~221170.  Only for HD~221170 are upper limits shown.  The bottom 
panel displays the difference defined as $\Delta$\eps{X} $\equiv$ 
\eps{X}$_{\rm obs}$ $-$ \eps{X}$_{\rm pred}$ for all four 
$r$-process-rich stars with respect to the scaled solar $r$-process 
predictions of \citet{jensim+04}.
}
\end{figure}

In Figure~\ref{fig.rmultistar}, we overplot the \citet{jensim+04}
scaled solar $r$-process predictions along with the abundances 
of \hd\ and other $r$-process-rich stars: \cs\ \citep{sneden+03}, 
HD~115444 \citep{westin+00}, and \bd\ \citep{cowan+02}.  Abundance 
studies of the ultra-$r$-process-rich metal-poor star \cs\ have shown 
that, while the abundances of $Z \ge$ 56 seem to match well the 
predictions from the scaled solar $r$-process, the lighter elements 
in the range of 40 $< Z <$ 56 are under-abundant \citep{sneden+03}.  
In general, the agreement in the abundances distribution patterns of 
the heavy $n$-capture elements ($Z \ge$ 56) in these $r$-process-rich 
stars suggests that a robust and perhaps even unique $r$-process 
produced these elements.  However, the same does not appear to be true 
of the lighter $n$-capture elements.

Observationally, many of the light $n$-capture elements are challenging 
to detect, requiring space-based observations.  With the aid of the 
Space Telescope Imaging Spectrograph (STIS) on the 
{\it Hubble Space Telescope} ($HST$), \citet{cowan+05} 
derived abundances of Ge, Zr, Os, and Pt in a sample of metal-poor 
stars, including \hd.  As noted in \S~\ref{anal-anal}, the spectrum of 
\hd\ is complex, and more so in the blue-violet-UV regions than in the 
green-yellow-red.  Despite the difficulties, their results for the 
abundances in common with this study are within the stated errors.

The differences in the light- versus heavy-$n$-capture elemental 
abundance patterns are not restricted to observations of metal-poor 
halo stars.  In an investigation of the abundances of short-lived 
isotopes in the early solar system (as inferred from meteoritic 
analyses) \citet{wbg96} found that distinctive SN sources or $r$-process 
sites were required to explain the incompatibility in the observed 
$^{129}$I/$^{182}$Hf ratio.  Both isotopes are short-lived and 
essentially of $r$-process origin, thus inferring two distinct 
$r$-process sites for the light isotopes below Ba, and for the heavy 
isotopes beyond Ba.  Their model of ``uniform production'' produced 
abundances that match those of actinides and the short-lived  
$^{182}$Hf, but overproduced the abundances of the lighter isotopes.  
This suggested to Wasserburg et al.\ that multiple $r$-process sites 
were required to produce both the heavy and light isotopes.  In 
another study of the short-lived radioactive isotopes in the early 
solar system, \citet{mc00} found that the abundances of the lighter 
isotopes {\em could} be explained by continuous galactic 
nucleosynthesis (providing a match to the inferred abundance of 
$^{129}$I, among other isotopes) combined with a relatively more recent 
injection of material into the condensing proto-solar nebula (providing 
a source for the relatively enhanced abundance of $^{182}$Hf, as well 
as $^{26}$Al, $^{36}$Cl, $^{41}$Ca, and $^{60}$Fe).  The $^{182}$Hf 
site suggested by \citet{mc00} is a ``fast $s$-process'' (with special 
conditions surrounding the subsequent mass lost from the progenitor 
star) rather than in the $r$-process proper. Production of $^{182}$Hf 
is essentially driven by neutron captures in the outer He-shell region 
by explosive nucleosynthesis \citep{meyer05}.  The Meyer \& Clayton 
model is reminiscent of a model discussed by \citet{ctc93}, who also 
invoked $s$-processing to explain some of the inferred abundances of 
isotopes normally considered to be of primarily $r$-process origins 
in the early solar system.

Although the \citet{ctc93}, \citet{wbg96} and \citet{mc00} studies 
disagree as to the proportions of ``uniform'' or ``continuous'' 
production required to explain the inferred meteoritic abundance 
patterns of the short-lived isotopes in the early solar system, all of 
the studies invoke another source, beyond continuous Galactic 
$r$-process production, to explain the discrepant abundances of isotopes 
normally considered to be of primarily $r$-process origin.   For further
discussion of this issue, we refer the reader to \citet{goswami+05} and
\citet{wasserburg+06} for recent reviews.

A phenomenological model for the formation of the $n$-capture abundance 
patterns observed in $r$-process-rich metal-poor stars has been 
developed by \citet{qw00} and \citet{wq00}.  The model replicates many
of the observed abundance patterns in very metal-poor stars with a mix 
of two different types of $r$-process events which occur on different
frequencies (high and low), and produce different sets of yields (heavy 
and light $n$-capture elements).  Another point of view taken to 
explain the light- versus heavy-$n$-capture abundance patterns has been 
to postulate the existence of both a ``main'' and a ``weak'' 
$r$-process.  The main $r$-process would produce the (under-abundant 
light-) $n$-capture abundance patterns observed in the low-metallicity 
$r$-process-rich stars.  A weak $r$-process, presumed to be a secondary 
process (\ie, dependent upon an existing heavy element distribution), 
and possibly the result of only a small neutron burst, would then 
operate on existing seeds to produce the observed solar abundance 
pattern in the light-$n$-capture elements (see \eg, Pfeiffer, Ott, \& 
Kratz 2001\nocite{pok01}; Thielemann et al.\ 
2001\nocite{thielemann+01}; Truran, Cowan, \& Fields 
2001\nocite{tcf01}).  Both of these views have been developed in 
response to the different abundance pattern characteristics between the 
light- and heavy-$n$-capture elements.

\begin{figure}
\epsscale{0.85}

\plotone{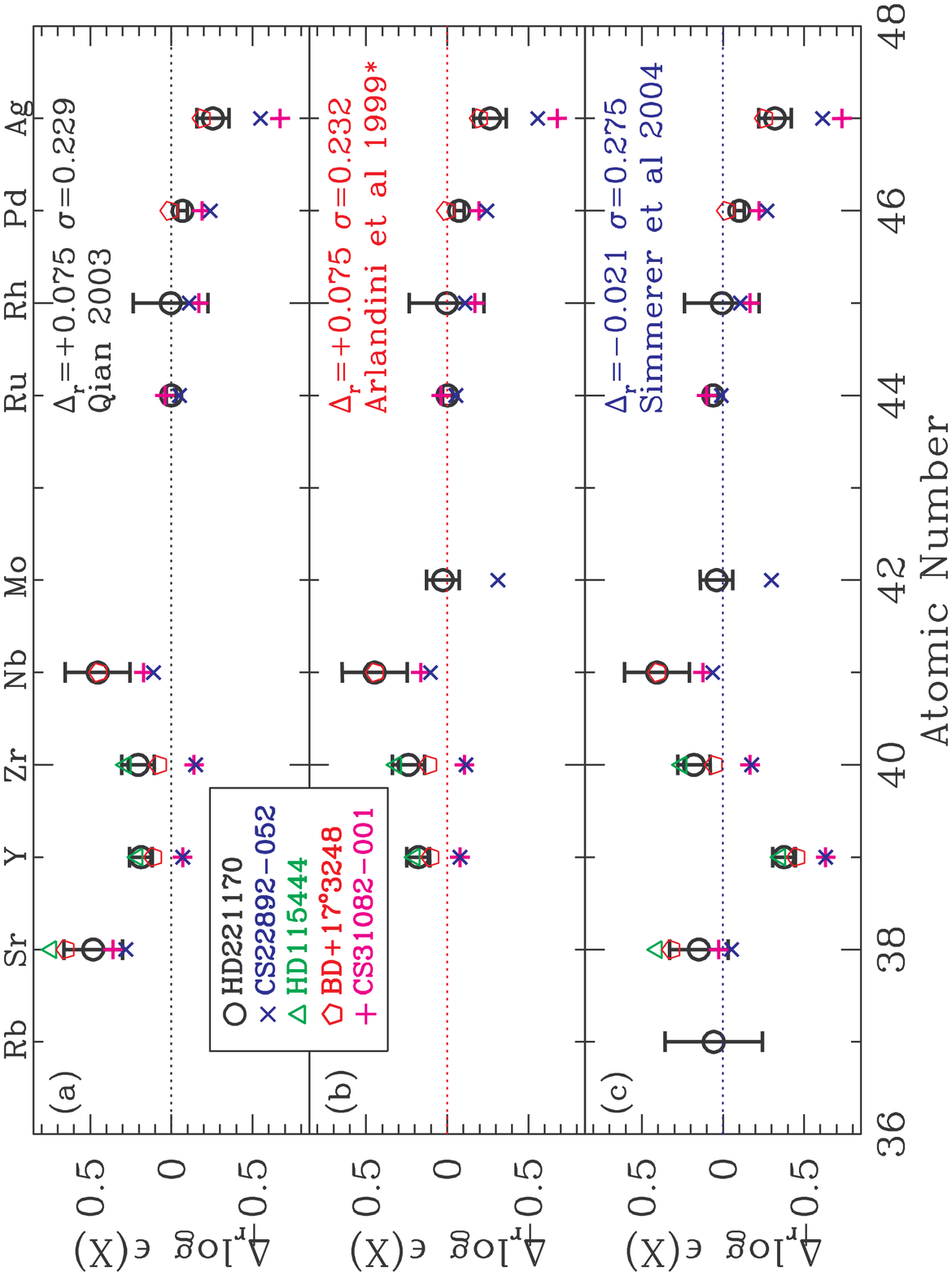}
\caption{
\label{fig.zoom} 
Comparison of the abundances of 37 $\le Z >$ 50 in \hd\ with those of 
CS~22892-052, HD~115444, BD+17\deg 3248, and the scaled-solar 
$r$-process predictions, with symbols and normalization as described 
in Figure~\ref{fig.rmultistar}.  Here, we also add the results for 
CS~31082-001 (magenta +; Hill et al.\ 2002).  The three panels display 
the difference ($\Delta$\eps{X} $\equiv$ \eps{X}$_{\rm obs}$ $-$ 
\eps{X}$_{\rm pred} = \Delta_{r}$) and scatter ($\sigma$) for all five 
$r$-process-rich stars with respect to the scaled-solar $r$-process 
predictions of \citet{qian03}, Arlandini* et al.\ (largely 1999; with 
modifications described in \S~\ref{comp-ss}), and \citet{jensim+04}.
}
\end{figure}

As shown in Figures~\ref{fig.riiiyu}--\ref{fig.rmultistar}, the 
abundances of $Z \ge$ 56 derived in this study are well-matched by the 
the scaled solar $r$-process pattern predictions.  Surprisingly, the 
abundances of \hd\ {\em also} seem to be a better match to the scaled 
solar $r$-process predictions in the 37 $ < Z <$ 48 elemental range 
than those of other $r$-process-rich stars.  Figure~\ref{fig.zoom} 
takes a closer perspective on this issue.  We display the abundance 
differences with respect to three predictions of the scaled solar 
$r$-process pattern: those of \citet{qian03}, Arlandini* et al.\ 
(largely 1999\nocite{arlandini+99}; with modifications as described in 
\S~\ref{comp-ss}) and \citet{jensim+04}).  The individual panels show 
the value of $<$$\Delta$(log~$\epsilon$)$>$ and $\sigma$ for the 
elements in common.  In addition to the $r$-process-rich stars shown 
in the previous figures, we have included here the abundances of 
CS~31082-001 \citep{hill+02}.  Regardless of whose scaled solar 
$r$-process predictions are referenced, the abundance patterns of 
$r$-process-rich stars in this light $n$-capture elemental range 
appear to have been stamped from a similar mould, but with an offset 
in the overall abundances.

In these $r$-process-rich metal-poor stars, it would appear that at 
least some of the light $n$-capture elements arise in sites that produce 
similar abundance yields, but with overall amounts that do not scale 
with the predicted scaled solar $r$-process values. Too little 
information exists in the literature regarding the abundance of Ga ($Z$ 
= 31) in metal-poor stars to warrant any comment beyond urging observers 
to obtain abundances of this element.  However, in the case of Ge ($Z$ = 
32), \citet{cowan+05} have shown that the abundance of [Ge/H] seems to 
track well the abundance of [Fe/H] whereas [Ge/Fe] appears to be 
independent of the [Eu/Fe]-ratio.   The behaviour of [Zr/Fe], on the 
other hand, is qualitatively different.  As can be seen in Figure 10 of 
the extensive investigation of Sr, Y, and Zr by \citet{travaglio+04}, 
[Zr/Fe] has only a mild dependence on the abundance of [Fe/H].  However, 
the largest [Zr/Fe] abundance displayed by Cowan et al.\ (their Figure 
6\nocite{cowan+05}) corresponds to the star with the largest [Eu/Fe] 
abundance, \cs.  Regarding other elements in this atomic mass range, the 
abundances of Pd and Ag have not been reported for HD~115444, nor have 
Mo, Ru and Rh for either it or for \bd.  

Iron-peak and transiron isotopes can be produced in the neutron-rich
$\alpha$-rich freeze-out process.  Such high entropy environments,
such those of nascent neutron stars, possess excess neutron and 
$\alpha$-particle distributions that can lead to the production of 
iron-peak isotopes, particularly those to the high-$Z$ side of the 
iron-peak (see \eg\ Arnett, Truran, \& Woosley 1971\nocite{atw71}; 
Woosley \& Hoffman 1992\nocite{wh92}; Nakamura et al.\ 
1999\nocite{nakamura+99}).  This process is not necessarily coupled to 
the $n$-capture nucleosynthesis of higher atomic mass isotopes.  Thus, 
it has been suggested that the $\alpha$-rich freeze-out process may be 
responsible for the behaviour of Ge abundances with metallicity 
\citep{cowan+05}.  Furthermore, the $\alpha$-rich freeze-out process
has been invoked as a possible explanation for the over-abundance of 
Sr \citep{ms99} and Y \citep{venn+04} observed in some metal-poor halo 
stars.  In the case of HD~221170, however, a boost from an overall 
increase in the efficiency of the $\alpha$-rich freeze-out does not 
successfully explain its light-$n$-capture abundance pattern.  The
abundances of lighter transiron elements (and sensitive to the neutron 
excess and the entropy; see \eg\ Arnett et al.\ 1971\nocite{atw71}) 
such as Cu ($Z$ = 29), Zn ($Z$ = 30), and Ge ($Z$ = 32), all have 
values in \hd\ that are in accord with those derived for \bd, 
HD~115444, and \cs.  Thus, the abundances of the elements in \hd\ in 
the range of 28 $< Z \le$ 32 do not seem to support an $\alpha$-rich 
freeze-out explanation for the overall relative enhancement in \hd\ 
of the abundances of elements in the range of 37 $ < Z <$ 48.  We 
refer the reader to \citet{bisterzo+04} for more extensive 
discussions regarding the complex origins of Cu and Zn, and to 
\citet{travaglio+04} for Sr, Y, and Zr.

The understudied abundances in the atomic mass range between the
iron-peak and $Z <$ 56 present a shopping list of elements which need 
to be examined and reported for a larger sample of metal-poor stars.  
Interestingly, while we can only report an upper limit for the 
abundance of Sn in \hd, the value is significantly lower than the 
predicted scaled solar value or of that derived in the spectrum of 
\cs .  While some light $n$-capture elements with features only in 
the UV wavelengths are limited to satellite observations, the 
near-UV-sensitive detector on Keck~I places at least some of the 
understudied elements within the reach of a ground-based facility 
(see Table~\ref{hd.analysis}).

\section{NUCLEOCOSMOCHRONOMETRY OF HD~221170 AND OTHER $r$-PROCESS-RICH STARS\label{age}}

With its very long half-life, the detection of Th in HD~221170 allows 
for nucleocosmochronometric estimates for the age of this star. The 
ideal chronometer pair for these age determinations is Th/U, since 
both elements are made entirely in the $r$-process and are nearby in 
mass number (see discussions in the reviews of Truran et al. 
2002\nocite{truran+02}, Sneden \& Cowan 2003\nocite{sc03}, Cowan \& 
Sneden 2004\nocite{cs04}, and Cowan \& Thielemann 2004\nocite{ct04}). 
Th/U has been used to determine the ages of CS~31082-001 (12.5 $\pm$ 
3 Gyr, Cayrel et al.\ 2001\nocite{cayrel+01}; 14 $\pm$ 2.4 Gyr, Hill 
et al.\ 2002\nocite{hill+02}; 15.5 $\pm$ 3.2 Gyr, Schatz et al.\ 
2002\nocite{schatz+02}; 14.1 $\pm$ 2.5 Gyr, Wanajo et al.\ 
2002\nocite{wanajo+02}) and \bd (13.8 $\pm$ 4 Gyr, Cowan et al.\ 
2002\nocite{cowan+02}). However, uranium (with its inherently low 
abundance and frequent spectral blending with strong atomic and 
molecular features) is difficult to detect in most stars and was not 
seen in HD~221170. 
 
Chronometric age estimates  based upon the Th/Eu ratio have been made 
for a number of stars (see \eg, Sneden et al.\ 1996\nocite{sneden+96}, 
2000a\nocite{sneden+00a}, 2003\nocite{sneden+03}; Cowan et al.\ 
1997\nocite{cowan+97}, 1999\nocite{cowan+99}; Pfeiffer, Kratz, \& 
Thielemann 1997\nocite{pkt97}; Westin et al.\ 2000\nocite{westin+00}; 
Johnson \& Bolte 2001\nocite{jb01}; del Peloso, da Silva, \& 
Arany-Prado 2005\nocite{dda05}).  Our abundance determinations for \hd\
indicate a ratio of \eps{Th/Eu} = $-$0.60 with a typical observational 
uncertainty of $\sigma$(log $\epsilon$) = 0.05.  This ratio is 
virtually identical to that found for \cs\ ($-$0.62, Sneden et al.\ 
2003\nocite{sneden+03}),  \bd\ ($-$0.51, Cowan et al.\ 
2002\nocite{cowan+02}) and HD~115444 ($-$0.60, Westin et al.\ 
2000\nocite{westin+00}).  The abundance ratio for HD~221170 is also 
consistent with the average \eps{Th/Eu} ratio found in two giant stars 
in the globular cluster M15 ($<$$-$0.62$>$, Sneden et al.\ 
2000b\nocite{sneden+00b}). 

As indicated earlier in this paper, our abundance results do not agree 
with those of  \citet{yu+05} -- we do not find an enhanced abundance 
of Th in HD~221170 and the (stable) $n$-capture elemental abundances 
in this star are consistent with the predicted scaled solar system 
$r$-process abundance distribution pattern and with other $r$-process 
rich stars (\eg, \cs, \bd, CS~31082-01).  Thus, in contrast to the 
suggestion of \citet{yu+05}, we find that the abundance ratio of Th/Eu 
is indeed usable as a chronometer in this star.

Chronometric age estimates depend sensitively upon the initial 
predicted values of Th/Eu, which in turn depend on the nuclear mass 
formulae and $r$-process models employed in making those 
determinations (see discussion in Schatz et al.\ 
2002\nocite{schatz+02} and Cowan \& Sneden 2004\nocite{cs04}). 
Calculations of the $r$-process are designed to replicate the solar 
system isotopic (and elemental) abundance distribution. The same 
calculations that reproduce the stable solar system (and stellar) 
elemental abundances are then used to predict the radioactive 
abundances and thus the Th/Eu ratio.  Comparing the observed Th/Eu 
abundance (0.25) in HD~221170 with the predicted $r$-process values 
can directly yield an age determination via the following
relation:
\begin{equation}
\left(\frac{\rm Th}{\rm Eu}\right)_{\star} =
 \left(\frac{\rm Th}{\rm Eu}\right)_{r} \exp[-t/\tau_{\rm Th}]
\label{eqn.1}
\end{equation}
where $\tau_{\rm Th}$ represents the characteristic decay
timescale of Th (20.27~Gyr) and $t$ the inferred age.

Earlier theoretical calculations predicted an initial value of Th/Eu = 
0.48 in an $r$-process site \citep{cowan+99}, while more recent 
calculations, constrained by some recent experimental data, suggest a 
value of 0.42 \citep{sneden+03}.  \citet{kratz+04} find Th/Eu = 0.42 
for traditional  (\ie, ``iron seed'') calculations and a value of 0.48 
for conditions that might be more typical of the high-entropy wind 
scenario in a Type II SN. Employing an average of these two initial 
abundance predictions for Th/Eu suggests an age for HD~221170 of 11.7 
$\pm$ 1.4 Gyr. The error uncertainty here is strictly from the two 
different theoretical calculations. Observational abundance errors 
(\eg, log $\epsilon$ = 0.05) would also contribute to the general age 
uncertainty.  Since these errors are uncorrelated, our best age 
estimate for HD~221170, based upon the Th/Eu chronometer, is 11.7 
$\pm$ 2.8 Gyr. 

If we put aside the theoretical considerations for a moment, and 
assume the \citet{ag89} value of Th/Eu = 0.463 in the early solar 
system, then for the Th/Eu of 0.2512 $\pm$ 0.05 in HD~221170, we can 
use this information to derive an inferred age based on the adopted 
Th/Eu in the early solar system.  Employing Equation~\ref{eqn.1} 
($0.25 \pm 0.05 = 0.463 \exp[-t/20.27~{\rm Gyr}]$), the inferred age 
is 12.4$^{+4.5}_{-3.7}$~Gyr, in agreement with that derived employing 
the theoretical $r$-process calculations.  

The age we derive for \hd\ is within the range of cosmic ages 
determined by the results of the Wilkinson Microwave Anisotropy Probe 
experiment (WMAP), both those combined with results from the Sloan 
Digital Sky Survey (14.1$^{+1.0}_{-1.9}$~Gyr; Tegmark et al.\ 
2004\nocite{tegmark+04}), as well as those combined with earlier cosmic 
microwave background (CMB) and large-scale structure data (13.7 $\pm$ 
0.2~Gyr; Spergel et al.\ 2003\nocite{spergel+03}).  These ages are also 
in agreement with the main sequence turn-off ages of the oldest globular 
clusters (12.5$^{+3.5}_{-2.4}$~Gyr; Krauss \& Chaboyer 
2003\nocite{kc03}; further refined to 12.6$^{+3.4}_{-2.2}$~Gyr with the 
inclusion of CMB data; Jimenez et al.\ 2003\nocite{jimenez+03}).  In 
turn, the white dwarf cooling curve age of 12.1 $\pm$ 0.9~Gyr, derived 
for globular cluster M4 (NGC~6121) by \citet{hansen+04}, is consistent 
with ages derived from the main sequence turn-off for this cluster.

Although our age estimate for HD~221170 is consistent with those 
reported for $r$-process-rich stars, additional experimental and 
theoretical  studies -- particularly of very neutron-rich nuclei -- are 
required to reduce meaningfully the chronometric age uncertainties.  In 
particular, employing certain other nuclear mass formulae would lead to 
a wider range of initial abundance ratios and correspondingly wider 
range in age estimates. We note, however, that some of the  mass models 
that have been employed to predict (wide-ranging) initial 
chronometric-age abundances do not reproduce the solar Pb and Bi 
abundances, or adequately predict the nuclear properties of nuclei far 
from stability (see Cowan et al.\ 1999\nocite{cowan+99} and Schatz et 
al.\ 2002\nocite{schatz+02}). 

The $r$-process-rich star CS~31082-001 is one well-documented case in
which Th/Eu is anomalously high.  While the abundances of the  stable 
elements (through the \third\ $r$-process peak) in CS~31082-001 are 
consistent with the predicted scaled solar system $r$-process 
distribution, Th and U are enhanced with respect to the other $n$-capture 
elements \citep{hill+02,schatz+02}.  Thus, in this star the Th/Eu 
chronometer leads to unreasonably low age determinations, while the Th/U 
ratio predicts an age consistent with other ultra metal-poor halo stars. 
We note, however, that the reported lead abundance in CS~31082-001
\citep{plez+05}, which predominantly results from $\alpha$-decay of Th 
and U isotopes, seems to be too low with respect to such high abundance 
values of these radioactive elements \citep{kratz+04}.  \citet{qw01}
have suggested that the U in CS~31082-001, along with the enhanced 
abundances of Os and Ir, may be the result of contamination from a 
nearby SNeII event, sufficiently nearby to have come from a binary
companion.  If the binary companion survived, it likely became a 
stellar-mass black hole.  \citet{schlegel03} undertook a survey of 
all available X-ray data acquired in the vicinity of CS~31082-001.  
Observations by the ROSAT All-Sky Survey \citet{voges+99} set an upper 
limit on the X-ray luminosity which fall short of the expected flux 
emanating from the region surrounding a stellar-mass black hole at the 
assumed distance of CS~31082-001.  However, XMM-Newton observations 
have been undertaken to uncover lower X-ray fluxes which may be being 
emanated from the system.  

Models of SNeII explosions have found that $r$-process nucleosynthesis
can take place in the wind generated by neutrinos flowing outwards from 
the surface of a hot newborn neutron star (see references given in 
\S~1 as well as Otsuki et al.\ 2000\nocite{otsuki+00} and 
additional references therein).  Sasaqui, Kajino, and Balantekin 
(2005\nocite{skb05}) have investigated the effect that a sudden 
cessation of neutrino fluxes (resulting from the creation of a black 
hole) can have on the $r$-process nucleosynthesis yields.  They find 
that not only are the resulting yields be quite sensitive to the 
neutrino cutoff effect in their model, but that the largest impact is 
on the abundance of the Th and U isotopes. In the case of CS~31082-001, 
their model with no $r$-process cutoff (and, by implication, no 
cessation of the neutrino fluxes through the formation of a black hole) 
provides the best match to the observations of Th and U.

Clearly, additional observational and theoretical studies of this star, 
and others with enhanced actinide abundances, such as the recently 
discovered $r$-process-rich Th-rich star CS~30306-132 \citep{honda+04}, 
will be needed to better understand and further refine the chronometric 
age determinations.

\section{SUMMARY AND CONCLUSIONS\label{conclude}}

Employing high resolution data acquired with the Keck Observatory 
HIRES, and supplemented by high resolution data gathered with the 
McDonald Observatory 2d-coud\'e spectrograph, we have derived the 
abundances of 57 species in the metal-poor red giant star HD~221170.
The stellar atmospheres have been derived employing spectroscopic
constraints -- eliminating abundance trends with the excitation 
potentials, equivalent widths (EW), and ionization states of 
$\sim$200 iron-peak lines -- and are in good agreement with the 
\teff\ and \logg\ expectations based on photometric and parallax 
measurements.

In the abundance analysis, single-source and recent $gf$-values were 
employed wherever possible in order to diminish uncertainties resulting 
from systematics in combining multiple sources.  Single, unblended 
features were analysed using EWs, and more complex features synthesized, 
with hyperfine structure (HFS) and isotopic splitting accounted for where 
necessary.  In total, over 700 features were analysed in this star, with 
the majority arising from neutron-capture ($n$-capture) transitions.  
Nearly all of the strongest $n$-capture transitions occur in the complex 
blue-UV spectral regions where blended features are the rule, not the 
exception.  Therefore, we employed full synthetic spectrum analyses of 
more than half of the $n$-capture features we investigated.  Included in 
our analysis are numerous $n$-capture features with newly-determined 
atomic parameters.  We provide improved wavelengths and full hyperfine 
and isotopic patterns for 24 lines of \nion{Eu}{ii}, and complete 
hyperfine patterns for 80 lines of \nion{La}{ii}.  More than 350 
transitions of 35 species yielded abundances for 30 $n$-capture elements 
and significant upper limits for three others.

The resulting $n$-capture abundance pattern distribution for $Z \ge$ 56
is fit well by the predicted scaled solar system $r$-process abundances,
as has been seen in other $r$-process-rich stars such as CS~22892-052, 
\bd, and HD~115444.  We derive ratios of \eps{Th/La} = $-$0.73 ($\sigma$ 
= 0.06) and \eps{Th/Eu} = $-$0.60 ($\sigma$ = 0.05), values in excellent 
agreement with those previously derived for other $r$-process-rich 
metal-poor stars, including the giant stars of globular cluster M15.  
Also comparable is the inferred age, based upon the Th/Eu chronometer, 
of 11.7 $\pm$ 2.8~Gyr (which includes the estimated uncertainty arising 
from both the measured abundances and the predicted Th/Eu production 
ratio) and in accord with the cosmic age derived from the measurements 
made by the Wilkinson Microwave Anisotropy Probe experiment.  Thus, the
abundance ratio of Th/Eu is indeed usable as a chronometer in this
star.

Surprisingly, in contrast to the abundance patterns of other
$r$-process-rich metal-poor stars, the abundances of \hd\ also seem to 
be a better match to the scaled solar $r$-process predictions in the 
lighter $n$-capture element range of 37 $ < Z <$ 48.  Based on the 
abundances of iron-peak and the lighter transiron elements, however, 
the source of the light $n$-capture agreement in HD~221170 does not 
appear to lie in the abundance predictions from the neutron-rich 
$\alpha$-rich freeze-out process.

\acknowledgments 

The successful efforts by the HIRES CCD upgrade group are acknowledged, 
as is the expertise of Keck staff during the run.  We are grateful for 
the privilege to observe on the revered summit of Mauna Kea.  III is 
indebted to John Norris and Sean Ryan for their guidance on FIGARO/IRAF 
\'echelle reductions; to Andy McWilliam, Iv\'an Ram\'irez, and Jorge 
Mel\'endez for supplying software tools; to Jon Fulbright and Andy 
McWilliam for sending linelists in electronic form; to Norbert 
Christlieb for sending results in electronic form; and to George 
Preston, Yong-Zhong Qian, Jim Truran, Gerry Wasserburg, and Dave
Yong for illuminating discussions; and to Wako Aoki, Bob Kraft, 
and Alexander Yushchenko for their careful reading of (and helpful 
comments on) an earlier version of this paper; and to the reviewers 
for useful suggestions incorporated into the manuscript.

This research was supported via funding to III through a 
Carnegie-Princeton Fellowship and from NASA through Hubble Fellowship 
grant HST-HF-01151.01-A from the Space Telescope Science Inst., 
operated by AURA, under NASA contract NAS5-26555; from the NSF 
through grants AST03-07495 to CS, AST-0506324 to JEL, and AST03-07279 
to JJC; and from the Italian MIUR-FIRB Project ``The Astrophysical 
Origin of the Heavy Elements beyond Fe" to RG and SB.  We appreciate 
the use of NASA's Astrophysics Data System Bibliographic Services; 
the Database on Rare Earths at Mons University; the SIMBAD database, 
operated at CDS, Strasbourg, France; data products from the Two 
Micron All Sky Survey, which is a joint project of the University of 
Massachusetts and the Infrared Processing and Analysis 
Center/California Institute of Technology, funded by NASA and the 
NSF; and NED, the NASA/IPAC Extragalactic Database which is 
operated by the Jet Propulsion Laboratory, California Institute of 
Technology, under contract with NASA.

\appendix

\section{Transition Data for $r$- and $s$-Process Surrogates}

Abundances of Eu and La are useful as surrogates for $r$- and 
$s$-process abundances in studies of Galactic halo stars (\eg, work on 
the rise of $s$-process by Simmerer et al.\ 2004\nocite{jensim+04}). 
Modern transition probabilities based on a combination of radiative 
lifetimes from laser induced fluorescence (LIF) measurements and 
branching fractions from an FTS for the important strong lines of 
\nion{Eu}{ii} and \nion{La}{ii} were reported by Lawler et al.\ 
(2001a\nocite{lawler+01a}, 2001c\nocite{lawler+01c}).  These authors 
also collected the best available isotope shift (IS) and HFS data for 
the ground, low metastable, and resonance levels of \nion{Eu}{ii} and 
\nion{La}{ii}.  They determined some of the missing IS and HFS data 
using FTS spectra.

\subsection{Updated La Transition Data}

In Table~\ref{hd.lahfs}, we present updated HFS patterns for 80
\nion{La}{ii} features, including 30 employed in this study.  
For all but four levels, these were computed using the best data 
compiled and/or measured by \citet{lawler+01a}: HFS constants 
for the the upper levels at 22106.02 and 22537.30 \invcm\ were 
updated using \citet{li+01}, and the upper levels at 24462.66 
and 25973.37 \invcm\ were updated using \citet{imura+03}. The 
Component (Comp.) positions are referenced to the center-of-gravity 
(COG) wavenumbers and wavelengths. Transition wavenumbers were 
computed from NIST energy levels \citep{martin+78} and converted to 
air wavelengths using the standard index of air (Edlen 
1953\nocite{edlen53}, 1966\nocite{edlen66}).  Comp.\ Strengths are 
normalized to sum to one.  

\subsection{Updated Eu Transition Data}

During our earlier work on \nion{Eu}{ii}, it became apparent that the 
energy levels could be improved beyond that available from NIST 
(Martin et al.\ 1978\nocite{martin+78}).   The NIST energy levels are 
from grating spectrometer measurements by \citet{russell+41}.   
Interferometric FTS accuracy ($\sim$ $\pm$ 0.003 \invcm) is at least 
an order of magnitude better than the accuracy of grating spectrometer 
measurements on photographic plates, particularly on lines with very 
wide ($\sim$1 \invcm) IS and HFS patterns.   Many of the critical 
\nion{Eu}{ii} lines to the ground and first metastable levels have 
very wide structure.  The wide structure is in part due to the large 
($\sim$0.150 \invcm) value of the IS from an unpaired 6s electron in 
the ground and lowest metastable levels.  

Astrophysical data on halo stars from modern large telescopes has now 
reached the level of quality that isotopic abundances of Eu can be 
determined (\eg, Sneden et al.\ 2002\nocite{sneden+02}).  Evidence to 
date supports a uniform isotopic mix from all $r$-process events.  
The lines of \nion{Eu}{ii} connected to ground and lowest metastable 
levels are ideal for such studies.

In Table~\ref{hd.tabN+1} we report FTS measurements of COG wavenumbers 
with a Solar System isotopic mix of $^{151}$Eu and $^{153}$Eu 
\citep{rt97} for selected levels of \nion{Eu}{ii}.  The procedure used 
in the \nion{Ho}{ii} work by \citet{lawler+04} was followed.  Internal 
standard \nion{Ar}{ii} and \nion{Ar}{i} lines were used to calibrate the 
scale of four FTS spectra.  The \nion{Ar}{ii} reference wavenumbers are 
from \citet{lt88} with a small correction by \citet{whaling+02}.  The 
\nion{Ar}{i} reference wavenumbers are from \citet{whaling+02}.  
Non-linear least-square fits of the complete line profiles, including 
all isotopic and hyperfine components, were performed to determine COG 
wavenumbers for 24 \nion{Eu}{ii} transitions.  A global least-square 
adjustment of the energy levels was performed using the redundant 
measured wavenumbers for the 24 transitions connected to the 13 levels 
of interest (12 excited levels, and the ground level defined as 0.00 
\invcm).
  
Since the publication of the earlier work on \nion{Eu}{ii} 
\citep{lawler+01c}, requests have been made for lists of the complete 
hyperfine and IS patterns of the important lines of \nion{Eu}{ii}.
While this information can be reconstructed from data in the earlier 
paper, some effort is required.   Here we provide the complete isotopic 
and HFS structure patterns along with improved COG transition 
wavelengths (in air) for a Solar System isotopic mix of $^{151}$Eu and 
$^{153}$Eu in Table~\ref{hd.tabN+2}.   Energy levels from this work were 
combined with the best HFS constants from Table 4, and recommended line 
isotope shifts from Table 5 of \citet{lawler+01c}.  Comp.\ Positions are 
referenced to the COG values and Comp.\ Strengths are normalized to 
1.000 for each line. The first half of the components listed for each 
line are from $^{151}$Eu and the second half are from $^{153}$Eu.

\clearpage

\begin{center}
\begin{deluxetable}{clrrrrlc}
\tabletypesize{\scriptsize}
\tablenum{1}
\tablewidth{0pt}
\tablecaption{Abundances Derived from Individual Features\label{hd.analysis}}
\tablecolumns{8}
\tablehead{
\colhead{(1)}                  &
\colhead{(2)}                  &
\colhead{(3)}                  &
\colhead{(4)}                  &
\colhead{(5)}                  &
\colhead{(6)}                  &
\colhead{(7)}                  &
\colhead{(8)}                  \\
\colhead{Ion}                  &
\colhead{$\lambda$}            &
\colhead{$\chi$}               &
\colhead{log}                  &
\colhead{EW}                   &
\colhead{\eps{X}}              &
\colhead{[X/Fe]}               &
\colhead{Notes}               \\
\colhead{}                     &
\colhead{(\AA)}                &
\colhead{(eV)}                 &
\colhead{{\it gf}-value}       &
\colhead{(m\AA)\tablenotemark{(a)}}&
\colhead{}                     &
\colhead{\tablenotemark{(b)}}  &
\colhead{}                      
}         
\startdata
\cutinhead{\eps{C}$_{\sun}$ = 8.56 ($Z$ = 6)\tablenotemark{(c)}}
 I &   CH-band  &\nodata 
	           &\nodata  & syn & 5.67 & $-$0.71 & \nodata \\
\cutinhead{\eps{N}$_{\sun}$ = 8.05 ($Z$ = 7)}
 I &   CN-band  &\nodata 
	           &\nodata  & syn & 6.46 &  $+$0.59 & \nodata \\
\cutinhead{\eps{O}$_{\sun}$ = 8.93 ($Z$ = 8)}
 I & 6300.31 & 0.00 & $-$9.750 &   syn & $+$6.97 & $+$0.22 & \nodata  \\
 I & 6363.79 & 0.02 &$-$10.250 &   syn & $+$6.97 & $+$0.22 & \nodata  \\
\cutinhead{\eps{Na}$_{\sun}$ = 6.33 ($Z$ = 11)}
 I & 5682.63 & 2.10 & $-$0.699 &   syn & $+$3.89 & $-$0.26 & \nodata \\
 I & 5688.00 & 2.10 & $-$0.456 &   syn & $+$3.95 & $-$0.20 & ~~HFS \\
 I & 5889.95 & 0.00 & $+$0.112 & 220.2 &\nodata & \nodata & \tablenotemark{(d)}~ \\
 I & 5895.92 & 0.00 & $-$0.191 & 212.3 &\nodata & \nodata & \tablenotemark{(d)}~ \\
\enddata
\tablecomments{These results are based on Keck HIRESb spectrum features for wavelengths in the range $\sim$3050 $\le \lambda \le$ 5895~\AA, and McDonald 2d-coud\'e spectrum features for redder wavelengths.  See \S~3.}
\tablenotetext{(a)}{In the place of an EW  measurement, ``syn'' denotes a feature for which our derived abundance relies on a spectrum synthesis computation.}
\tablenotetext{(b)}{The abundances derived for individual feature relative to the scaled solar value:[\nion{Fe}{i}/H], [\nion{Fe}{ii}/H], or [X/$<$Fe$>$] for $<$[Fe/H]$>$ = --2.18.  Upper limits are denoted :UL.}
\tablenotetext{(c)}{Derived $^{12}$C/$^{13}$C ratio = 7 $\pm$ 2.}
\tablenotetext{(d)}{Excluded from analysis (non-negligible non-LTE correction required.).}
\tablenotetext{(e)}{Hyperfine structure not considered for EW $<$ 30m\AA.}
\tablenotetext{(f)}{Excluded from analysis (HFS information unavailable for non-negligle HFS correction).}
\end{deluxetable}
\tablecomments{Displayed here is a portion (of the entire electronically-available abundance table) as a guide to the general form and content.}

\end{center}

\begin{center}
\begin{deluxetable}{clcc}
\tabletypesize{\scriptsize}
\tablenum{2}
\tablewidth{0pt}
\tablecaption{\nion{Sc}{ii} and \nion{Mn}{i} Linelists\label{hd.scmn}}
\tablecolumns{4}
\tablehead{
\colhead{$\lambda$}           &
\colhead{Species}             &
\colhead{$\chi$}              &
\colhead{$\log{}$\tablenotemark{(a)}}            \\
\colhead{(\AA)}               &
\colhead{}                    &
\colhead{(eV)}                &
\colhead{$gf$-value}
}
\startdata
  4400.379 & \nion{Sc}{ii} & 0.605 &   -2.019 \\
  4400.383 & \nion{Sc}{ii} & 0.605 &   -1.821 \\
  4400.383 & \nion{Sc}{ii} & 0.605 &   -1.196 \\
  4400.387 & \nion{Sc}{ii} & 0.605 &   -1.767 \\
  4400.387 & \nion{Sc}{ii} & 0.605 &   -1.429 \\
\nodata
\enddata   
\tablenotetext{(a)}{Hyperfine structure information 
adopted from \citet{kb95}, with \loggf's normalized
to those adopted in this study (see \S~\ref{anal-ew}).}
\tablecomments{Displayed here is a portion (of the entire electronically-available Sc and Mn linelist) as a guide to the general form and content.}
\end{deluxetable}   
\end{center}

\begin{center}
\begin{deluxetable}{crc|rrccrl}
\tabletypesize{\scriptsize}
\tablenum{3}
\tablewidth{0pt}
\tablecaption{\hd22: Abundance Summary\label{hd.abtot}}
\tablecolumns{9}
\tablehead{
\colhead{Species}        &
\colhead{$Z$}            &
\colhead{\eps{X}}        &
\colhead{\eps{X}}        &
\colhead{$\pm$}          &
\colhead{$\sigma$}       &
\colhead{$\sigma$}       &
\colhead{N}              &
\colhead{[X/Fe]}         \\
\colhead{}               &
\colhead{}               &
\colhead{$\sun$}         &
\colhead{$\star$}        &
\colhead{}               &
\colhead{(lines)}        &
\colhead{(adopted)}      &
\colhead{}               &
\colhead{\tablenotemark{(a)}} 
}         
\startdata
$<$\nion{C}{i}$>$   &  6    &8.56   &    5.67 &\nodata&\nodata&    0.30&   CH & $-$0.71 \\
$<$\nion{N}{i}$>$   &  7    &8.05   &    6.46 &\nodata&\nodata&    0.35&   CN & $+$0.59 \\
$<$\nion{O}{i}$>$   &  8    &8.93   &    6.97 &0.00   &   0.00&    0.10&   2 &$+$0.22 \\
$<$\nion{Na}{i}$>$  & 11    &6.33   &    3.92 &0.03   &   0.04&    0.10&   2 &$-$0.23 \\
$<$\nion{Mg}{i}$>$  & 12    &7.58   &    5.81 &0.03   &   0.06&    0.06&   6 &$+$0.41 \\
$<$\nion{Si}{i}$>$  & 14    &7.55   &    5.85 &0.05   &   0.09&    0.09&   4 &$+$0.48 \\
$<$\nion{Ca}{i}$>$  & 20    &6.36   &    4.53 &0.02   &   0.09&    0.09&  15 &$+$0.35 \\
$<$\nion{Sc}{ii}$>$ & 21    &3.10   &    1.01 &0.02   &   0.06&    0.06&   9 &$+$0.09 \\
\nion{Ti}{i}        &\nodata&\nodata&    3.04 &0.03   &   0.12&    0.12&  13 &$+$0.23 \\
\nion{Ti}{ii}       &\nodata&\nodata&    3.16 &0.03   &   0.11&    0.11&  15 &$+$0.36 \\
$<$Ti$>$            & 22    &4.99   &    3.10 &0.02   &   0.13&    0.13&  28 &$+$0.30 \\
\nion{V}{i}         &\nodata&\nodata&    1.83 &0.03   &   0.07&    0.07&   6 &$+$0.01 \\
\nion{V}{ii}        &\nodata&\nodata&    2.11 &\nodata&\nodata&\nodata&\nodata & $+$0.01:UL \\
$<$V$>$             & 23    &4.00   &    1.83 &0.03   &   0.07&    0.07&   6 &$+$0.01 \\
\nion{Cr}{i}        &\nodata&\nodata&    3.31 &0.02   &   0.04&    0.04&   7 &$-$0.18 \\
\nion{Cr}{ii}       &\nodata&\nodata&    3.63 &0.04   &   0.09&    0.09&   7 &$+$0.15 \\
$<$Cr$>$            & 24    &5.67   &    3.47 &0.05   &   0.18&    0.18&  14 &$-$0.01 \\
\nion{Mn}{i}        &\nodata&\nodata&    2.91 &0.04   &   0.10&    0.10&   6 &$-$0.30 \\
\nion{Mn}{ii}       &\nodata&\nodata&    3.19 &\nodata&\nodata&\nodata&\nodata& $-$0.02:UL \\
$<$Mn$>$            & 25    &5.39   &    2.91 &0.04   &   0.10&    0.10&   6 &$-$0.30 \\
\nion{Fe}{i}        &\nodata&\nodata&    5.34 &0.01   &   0.12&    0.12& 157 &$-$2.18 \\
\nion{Fe}{ii}       &\nodata&\nodata&    5.32 &0.02   &   0.10&    0.10&  19 &$-$2.20 \\
$<$Fe$>$            & 26    &7.52   &    5.34 &0.01   &   0.12&    0.12& 176 &$-$2.18 \\
$<$\nion{Co}{i}$>$  & 27    &4.92   &    2.87 &0.04   &   0.12&    0.12&   9 &$+$0.13 \\
$<$\nion{Ni}{i}$>$  & 28    &6.25   &    4.16 &0.05   &   0.13&    0.13&   6 &$+$0.09 \\
$<$\nion{Cu}{i}$>$  & 29    &4.21   &    1.27 &\nodata&\nodata&    0.20&   1 &$-$0.76 \\
$<$\nion{Zn}{i}$>$  & 30    &4.60   &    2.51 &0.00   &   0.00&    0.10&   2 &$+$0.09 \\
$<$\nion{Ga}{i}$>$  & 31    &2.88   &    0.59 &\nodata&\nodata&    0.20&   1 &$-$0.54 \\
$<$\nion{Rb}{i}$>$  & 37    &2.60   &    0.70 &\nodata&\nodata&    0.30&   1 &$+$0.28 \\
\nion{Sr}{i}        &\nodata&\nodata&    0.48 &\nodata&\nodata&    0.20&   1 &$-$0.24 \\
\nion{Sr}{ii}       &\nodata&\nodata&    0.85 &0.04   &   0.08&    0.08&   4 &$+$0.13 \\
$<$Sr$>$            & 38    &2.90   &    0.74 &0.08   &   0.18&    0.18&   5 &$+$0.02 \\
$<$\nion{Y}{ii}$>$  & 39    &2.24   & $-$0.08 &0.02   &   0.07&    0.07&  16 &$-$0.13 \\
\nion{Zr}{i}        &\nodata&\nodata&    0.65 &0.06   &   0.13&    0.13&   5 &$+$0.23 \\
\nion{Zr}{ii}       &\nodata&\nodata&    0.68 &0.03   &   0.09&    0.09&  12 &$+$0.26 \\
$<$Zr$>$            & 40    &2.60   &    0.67 &0.02   &   0.10&    0.10&  17 &$+$0.25 \\
$<$\nion{Nb}{ii}$>$ & 41    &1.42   & $-$0.37 &\nodata&\nodata&    0.30&   1 &$+$0.39 \\
$<$\nion{Mo}{i}$>$  & 42    &1.92   &    0.03 &0.03   &   0.04&    0.10&   2 &$+$0.29 \\
$<$\nion{Ru}{i}$>$  & 44    &1.84   &    0.22 &0.01   &   0.03&    0.05&   6 &$+$0.56 \\
$<$\nion{Rh}{i}$>$  & 45    &1.12   & $-$0.35 &0.13   &   0.23&    0.23&   3 &$+$0.71 \\
$<$\nion{Pd}{i}$>$  & 46    &1.69   & $-$0.03 &0.02   &   0.03&    0.05&   3 &$+$0.46 \\
$<$\nion{Ag}{i}$>$  & 47    &1.24   & $-$0.50 &0.05   &   0.07&    0.10&   2 &$+$0.44 \\
$<$\nion{Sn}{i}$>$  & 50    &2.00   & $-$0.40 &\nodata&\nodata&\nodata&\nodata&$-$0.22:UL \\
$<$\nion{Ba}{ii}$>$ & 56    &2.13   &    0.21 &0.04   &   0.12&    0.12&   8 &$+$0.26 \\
$<$\nion{La}{ii}$>$ & 57    &1.14   & $-$0.73 &0.01   &   0.06&    0.06&  36 &$+$0.32 \\
$<$\nion{Ce}{ii}$>$ & 58    &1.55   & $-$0.41 &0.02   &   0.12&    0.12&  44 &$+$0.22 \\
$<$\nion{Pr}{ii}$>$ & 59    &0.71   & $-$1.04 &0.01   &   0.06&    0.06&  21 &$+$0.43 \\
$<$\nion{Nd}{ii}$>$ & 60    &1.45   & $-$0.35 &0.01   &   0.08&    0.08&  63 &$+$0.39 \\
$<$\nion{Sm}{ii}$>$ & 62    &1.00   & $-$0.66 &0.01   &   0.07&    0.07&  28 &$+$0.52 \\
$<$\nion{Eu}{ii}$>$ & 63    &0.51   & $-$0.86 &0.02   &   0.07&    0.07&  16 &$+$0.80 \\
$<$\nion{Gd}{ii}$>$ & 64    &1.12   & $-$0.46 &0.04   &   0.14&    0.14&  11 &$+$0.60 \\
$<$\nion{Tb}{ii}$>$ & 65    &0.28   & $-$1.21 &0.03   &   0.08&    0.08&   8 &$+$0.69 \\
$<$\nion{Dy}{ii}$>$ & 66    &1.10   & $-$0.32 &0.02   &   0.07&    0.07&  14  &+0.76 \\
$<$\nion{Ho}{ii}$>$ & 67    &0.51   & $-$0.97 &0.02   &   0.07&    0.07&   8  &+0.70 \\
$<$\nion{Er}{ii}$>$ & 68    &0.93   & $-$0.48 &0.06   &   0.13&    0.13&   4  &+0.77 \\
$<$\nion{Tm}{ii}$>$ & 69    &0.13   & $-$1.38 &0.02   &   0.06&    0.06&   6  &+0.68 \\
$<$\nion{Yb}{ii}$>$ & 70    &1.08   & $-$0.51 &\nodata&\nodata&    0.20&   1  &+0.59 \\
$<$\nion{Lu}{ii}$>$ & 71    &0.12   & $-$1.40 &\nodata&\nodata&\nodata&\nodata&+0.66:UL \\
$<$\nion{Hf}{ii}$>$ & 72    &0.88   & $-$0.94 &0.02   &   0.04&    0.04&   5  &+0.47 \\
$<$\nion{W}{i}$>$   & 74    &0.68   & $-$0.60 &\nodata&\nodata&\nodata&\nodata&+0.90:UL \\
$<$\nion{Os}{i}$>$  & 76    &1.45   &    0.16 &0.04   &   0.10&    0.10&   7  &+0.90 \\
$<$\nion{Ir}{i}$>$  & 77    &1.35   &    0.02 &0.09   &   0.13&    0.13&   2  &+0.85 \\
$<$\nion{Pb}{i}$>$  & 82    &1.85   & $-$0.09 &0.15   &   0.21&    0.21&   2  &+0.24 \\
$<$\nion{Th}{ii}$>$ & 90    &0.12   & $-$1.46 &0.02   &   0.05&    0.05&   7  &+0.60 \\
\enddata
\tablenotetext{(a)}{The abundances of individual ions relative to the scaled solar value: [\nion{Fe}{i}/H], [\nion{Fe}{ii}/H], or [X/$<$Fe$>$] for $<$[Fe/H]$>$ = $-$2.18.  Upper limits are denoted :UL.}
\end{deluxetable}
\end{center}

\begin{center}
\begin{deluxetable}{rrccrrr}
\tabletypesize{\scriptsize}
\tablenum{1}
\tablewidth{0pt}
\tablecaption{Isotopic Hyperfine Structure Patterns: $^{139}$La\label{hd.lahfs}}
\tablecolumns{7}
\tablehead{
\colhead{}                        &
\colhead{}                        &
\colhead{}                        &
\colhead{}                        &
\colhead{Comp.}                   &
\colhead{Comp.}                   &
\colhead{}                        \\
\colhead{Wavenumber}              &
\colhead{$\lambda$}               &
\colhead{$F$}                     &
\colhead{$F$}                     &
\colhead{Position}                &
\colhead{Position}                &
\colhead{Strength}                \\
\colhead{(\invcm)}                &
\colhead{(\AA)}                   &
\colhead{upper}             &
\colhead{lower}             &
\colhead{(\invcm)}                &
\colhead{(\AA)}                   &
\colhead{}                        
}         
\startdata
27549.30 & 3628.822 & 7.5 & 6.5 &  0.02310 & --0.003043 & 0.22222 \\
27549.30 & 3628.822 & 6.5 & 6.5 & --0.00840 &  0.001106 & 0.02618 \\
27549.30 & 3628.822 & 6.5 & 5.5 &  0.01396 & --0.001839 & 0.16827 \\
27549.30 & 3628.822 & 5.5 & 6.5 & --0.03570 &  0.004703 & 0.00160 \\
27549.30 & 3628.822 & 5.5 & 5.5 & --0.01334 &  0.001757 & 0.04196 \\
\nodata
\enddata
\tablecomments{Displayed here is a portion (of the entire electronically-available table of La HFS pattern values) as a guide to the general form and content.}
\end{deluxetable}
\end{center}

\begin{center}
\begin{deluxetable}{rrr}
\tabletypesize{\scriptsize}
\tablenum{2}
\tablewidth{0pt}
\tablecaption{\nion{Eu}{ii} Energy Levels\label{hd.tabN+1}}
\tablecolumns{3}
\tablehead{
\multicolumn{2}{c}{Energy Levels} &
\colhead{$J$}                     \\
\multicolumn{2}{c}{(\invcm)}      &
\colhead{}                        \\
\colhead{This Study\tablenotemark{(a)}}  &
\colhead{NIST}                    &
\colhead{}                        
}         
\startdata
          0.000   &       0.00   &  4 \\
       1669.261   &    1669.21   &  3 \\
       9923.046   &    9923.00   &  2 \\
      10081.730   &   10081.65   &  3 \\
      10312.869   &   10312.82   &  4 \\
      10643.584   &   10643.48   &  5 \\
      11128.440   &   11128.22   &  6 \\
      23774.372   &   23774.28   &  3 \\
      24207.962   &   24207.86   &  4 \\
      26172.959   &   26172.83   &  5 \\
      26838.574   &   26838.50   &  4 \\
      27104.177   &   27104.07   &  3 \\
      27256.437   &   27256.35   &  2 \\
\enddata
\tablenotetext{(a)}{Accurate to 0.003 \invcm.}
\end{deluxetable}
\end{center}

\begin{center}
\begin{deluxetable}{rrrrrrr}
\tabletypesize{\scriptsize}
\tablenum{3}
\tablewidth{0pt}
\tablecaption{Isotopic Hyperfine Structure Patterns: $^{151}$Eu and $^{153}$Eu Solar System Mix\label{hd.tabN+2}}
\tablecolumns{7}
\tablehead{
\colhead{}                        &
\colhead{}                        &
\colhead{}                        &
\colhead{}                        &
\colhead{Comp.}                   &
\colhead{Comp.}                   &
\colhead{}                        \\
\colhead{Wavenumber}              &
\colhead{$\lambda$}               &
\colhead{$F$}                     &
\colhead{$F$}                     &
\colhead{Position}                &
\colhead{Position}                &
\colhead{Strength}                \\
\colhead{(\invcm)}                &
\colhead{(\AA)}                   &
\colhead{upper}             &
\colhead{lower}             &
\colhead{(\invcm)}                &
\colhead{(\AA)}                   &
\colhead{}                        
}         
\startdata
27104.177 & 3688.4183 & 5.5 & 6.5 & --0.47314 & 0.064390 & 0.12395 \\
27104.177 & 3688.4183 & 5.5 & 5.5 & --0.13900 & 0.018916 & 0.01207 \\
27104.177 & 3688.4183 & 5.5 & 4.5 &  0.14364 & --0.019547 & 0.00057 \\
27104.177 & 3688.4183 & 4.5 & 5.5 & --0.10598 & 0.014423 & 0.09417 \\
27104.177 & 3688.4183 & 4.5 & 4.5 &  0.17665 & --0.024039 & 0.01840 \\
27104.177 & 3688.4183 & 4.5 & 3.5 &  0.40782 & --0.055498 & 0.00126 \\
\nodata
\enddata
\tablecomments{Displayed here is a portion (of the entire electronically-available table of Eu HFS pattern values) as a guide to the general form and content.}
\end{deluxetable}
\end{center}


\begin{thebibliography}{}


\bibitem[Alonso, Arribas, \& Mart\'inez(1996)]{aam96}
Alonso, A., Arribas, S., \& Mart\'inez-Roger, C.\ 1996, \aap, 313, 873

\bibitem[Alonso, Arribas, \& Mart\'inez(1999)]{aam99}
Alonso, A., Arribas, S., \& Mart\'inez-Roger, C.\ 1999, \aaps, 140, 261

\bibitem[Anders \& Grevesse(1989)]{ag89} 
Anders, E.\ \& Grevesse, N.\ 1989, Geochim.\ Cosmochim.\ Acta, 53, 197

\bibitem[Andersen et al.(1976)]{andersen+76}
Andersen, T., Petersen, P., \& Hauge, O.\ 1976, \solphys, 49, 211

\bibitem[Aoki et al.(2003)]{aoki+03}
Aoki, W., Honda, S., Beers, T.~C., \& Sneden, C.\ 2003, \apj, 586, 506

\bibitem[Aoki et al.(2003)]{aoki+02}
Aoki, W., Ryan, S.\ G., Norris, J.\ E., Beers, T.\ C., Ando, H.\ \& Tsangarides, S.\ 2002, \apj, 580, 1149

\bibitem[Argast et al.(2004)]{argast+04}
Argast, D., Samland, M., Thielemann, F.-K., \& Qian, Y.-Z.\ 2004, \aap, 416, 997

\bibitem[Arlandini et al.(1999)]{arlandini+99}
Arlandini, C., K\"appeler, F., Wisshak, K., Gallino, R., Lugaro, M., Busso, M., \& Straniero, O.\ 1999, \apj, 525, 886

\bibitem[Arnett et al.(1971)]{atw71}
Arnett, D., Truran, J.\ W.\ \& Woosley, S.\ E.\ 1971, \apj, 165, 87

\bibitem[Arnould \& Goriely(2001)]{ag01}
Arnould, M.\ \& Goriely, S.\ 2001, in Astrophysical Ages and Time Scales, (ed.) T.\ von Hippel, N.\ Manset, \& C.\ Simpson, ASP 245, 252

\bibitem[Barklem et al.(2005)]{barklem+05}
Barklem, P.\ S., Christlieb, N., Beers, T.\ C., Hill, V., Bessell, M.\ S., Holmberg, J., Marsteller, M., Rossi, S., Zickgraf, F.-J.\ \& Reimers, D.\ 2005, \aap, 439, 129

\bibitem[Bergstr\"om et al.(1988)]{bergstrom+88}
Bergstr\"om, H., Lundberg, H., Persson, A., \& Bi\'emont, E.\ 1988, \aap, 192, 335

\bibitem[Bernath et al.(1991)]{bernath+01}
Bernath, P.\ F., Brazier, C.\ R., Olsen, T., Hailey, R., \& Fernando, W.\ T.\ M.\ L.\ 1991, J.\ Mol.\ Spec., 147, 16

\bibitem[Bi\'emont et al.(2000)]{biemont+00}
Bi\'emont, E., Garnir, H.\ P., Palmeri, P., Li, Z.\ S., \& Svanberg, S.\ 2000, \mnras, 312, 116

\bibitem[Bi\'emont et al.(1981)]{biemont+81}
Bi\'emont, E., Grevesse, N., Hannaford, P., \& Lowe, R.\ M.\ 1981, \apj, 248, 867

\bibitem[Bi\'emont et al.(1982)]{biemont+82}
Bi\'emont, E., Grevesse, N., Kwiatkowski, M., \& Zimmermann, P.\ 1982, \aap, 108, 127

\bibitem[Bi\'emont et al.(2002)]{biemont+02}
Bi\'emont, E., Quinet, P., Dai, Z., Zhankui, J., Zhiguo, Z., Xu, H.\ \& Svanberg, S.\ 2002, J.\ Phys.\ B, 35, 4743

\bibitem[Bisterzo et al.(2004)]{bisterzo+04}
Bisterzo, S., Gallino, R., Pignatari, M., Pompeia, L., Cunha, K.\ \& Smith, V.\ 2004, Mem.\ S.\ A.\ It., 75, 741

\bibitem[Burris et al.(2000)]{burris+00} 
Burris, D.\ L., Pilachowski, C.\ A., Armandroff, T.\ A., Sneden, C., Cowan, J.\ J.\ \& Roe, H.\ 2000, \apj, 544, 302

\bibitem[Burstein \& Heiles(1982)]{bh82} 
Burstein, D.\ \& Heiles, C.\ 1982, \aj, 87, 1165

\bibitem[Butcher(1987)]{butcher87}
Butcher, H.\ R.\ 1987, Nature, 328, 127

\bibitem[Cameron(2003)]{cameron03} 
Cameron, A.\ G.\ W.\ 2003, \apj, 587, 327

\bibitem[Cameron et al.(1993)]{ctc93}
Cameron, A.\ G.\ W., Thielemann, F.-K., \& Cowan, J.\ J.\ 1993, Phys.\ Reports, 227, 283

\bibitem[Carney et al.(2003)]{carney+03}
Carney, B.\ W., Latham, D.\ W., Stefanik, R.\ P., Laird, J.\ B., \& Morse, J.\ A.\ 2003, \aj, 125, 293

\bibitem[Castelli \& Kurucz(2004)]{ck04} 
Castelli, F., \& Kurucz, R.\ L.\ 2004, Modelling of Stellar Atmospheres, N.\ E.\ Piskunov, W.\ W.\ Weiss, D.\ F.\ Gray, ASP 210, 20

\bibitem[Cayrel et al.(2001)]{cayrel+01}
Cayrel, R., et al.\ 2001,  Nature,  409, 691

\bibitem[Cohen et al.(2003)]{cohen+03}
Cohen, J.\ G., Christlieb, N., Qian, Y.-Z., \& Wasserburg, G.\ J.\ 2003, \apj, 588, 1082

\bibitem[Cowan et al.(1997)]{cowan+97}
Cowan, J.\ J.,  McWilliam, A.,  Sneden, C., \&  Burris, D.\ L.\  1997, \apj, 480, 246

\bibitem[Cowan et al.(1999)]{cowan+99}
Cowan, J.\ J., Pfeiffer, B., Kratz, K.-L., Thielemann, F.-K., Sneden, C., Burles, S., Tytler, D., \& Beers, T.\ C.\ 1999, \apj, 521, 194

\bibitem[Cowan \& Sneden(2004)]{cs04}
Cowan, J.\ J., \& Sneden, C.\ 2004, in Carnegie Observatories Astrophysics Series, Vol. 4: Origin and Evolution of the Elements, (ed.) A.\ McWilliam \& M. Rauch (Cambridge: Cambridge Univ.\ Press), 27 

\bibitem[Cowan \& Thielemann(2004)]{ct04}
Cowan, J.\ J., \& Thielemann, F.-K., 2004, Phys.\ Today, 57, 47 

\bibitem[Cowan et al.(2002)]{cowan+02}
Cowan, J.\ J., et al.\ 2002, \apj, 572, 861

\bibitem[Cowan et al.(2005)]{cowan+05} 
Cowan, J.\ J., et al.\ 2005, \apj, 627, 238

\bibitem[Cunningham \& Link(1967)]{cl67}
Cunningham, P.\ T., \& Link, J.\ K. 1967, J.\ Opt.\ Soc.\ Am., 57, 1000

\bibitem[del Pelosa et al.(2005)]{dda05}
del Peloso, E.\ F., da Silva, L., \& Arany-Prado, L.\ I.\ 2005, \aap, 434, 301

\bibitem[Den Hartog et al.(1987)]{denhartog+87}
Den Hartog, E.\ A., Duquette, D.\ W., \& Lawler, J.\ E.\ 1987, J.\ Opt.\ Soc.\ Am.\ B, 4, 48

\bibitem[Den Hartog et al.(2003)]{denhartog+03}
Den Hartog, E.\ A., Lawler, J.\ E., Sneden, C., \& Cowan, J.\ J.\ 2003, \apjs, 148, 543

\bibitem[Dinneen et al.(1991)]{dinneen+91}
Dinneen, T.\ P., Berrah Mansour, N., Kurtz, C., \& Young, L.\ 1991, \pra, 43, 4824
                        
\bibitem[Duquette \& Lawler(1985)]{dl85}
Duquette, D.\ W., \& Lawler, J.\ E.\ 1985, J.\ Opt.\ Soc.\ Am.\ B, 1, 1948

\bibitem[Edlen(1953)]{edlen53}
Edlen, B.\ 1953, J.\ Opt.\ Soc.\ Am., 43, 339

\bibitem[Edlen(1966)]{edlen66}
Edlen, B.\ 1966,  Metrologia, 2, 71

\bibitem[Erspamer \& North(2003)]{en03}
Erspamer, D.\ \& North, P.\ 2003, \aap, 398, 1121

\bibitem[ESA(1997)]{ESA97} 
ESA 1997, The Hipparcos and Tycho Catalogues (ESA SP-1200; Noordjwik: ESA)

\bibitem[Fedchak et al.(2000)]{fedchak+00}
Fedchak, J.\ A., Den Hartog, E.\ A., Lawler, J.\ E., Palmeri, P., Quinet, P., \& Bi\'emont, E.\ 2000, \apj, 542, 1109

\bibitem[Fitzpatrick \& Sneden(1987)]{fs87} 
Fitzpatrick, M.\ J., \& Sneden, C.\ 1987, \baas, 19, 1129

\bibitem[Fran\c{c}ois et al.(1993)]{fss93}
Fran\c{c}ois, P.,  Spite, M.\ \& Spite, F.\ 1993, \aap, 274, 821

\bibitem[Freiburghaus, Rosswog, \& Thielemann]{frt99} 
Freiburghaus, C., Rosswog, S.\ \& Thielemann, F.-K.\ 1999, \apj, 525, L121

\bibitem[Fuhr \& Wiese(2005)]{fw05}
Fuhr, J.\ R., \& Wiese, W.\ L.\ 2005, NIST Atomic Transition Probability Tables, CRC Handbook of Chemistry and Physics, (ed.) D.\ R.\ Lide, (CRC Press, Boca Raton, FL), 78, 10

\bibitem[Fulbright(2000)]{fulbright00} 
Fulbright, J.\ P.\ 2000, \aj, 120, 1841

\bibitem[Gilroy et al.(1988)]{gilroy+88}
Gilroy, K.\ K., Sneden, C., Pilachowski, C.\ A., \& Cowan, J.\ J.\ 1988, \apj, 327, 298

\bibitem[Gopka et al.(2004)]{gopka+04} 
Gopka, V.\ F., Yushchenko, A.\ V., Mishenina, T.\ V., Kim, C., Musaev, F.\ A., \& Bondar, A.\ V.\ 2004, Astron.\ Reports, trans.\ from 2004, Astron.\ Zhurnal, 48, 7

\bibitem[Goswami et al.(2005)]{goswami+05}
Goswami, J.\ N., Marhas, K.\ K., Chaussidon, M., Gounelle, M.\ \& Meyer, B.\ S.\ 2005, in Chondrites and the Protoplanetary Disk, (ed.) A.\ N.\ Krot, E.\ R.\ D.\ Scott, \& B.\ Reipurth, ASP 341, 485

\bibitem[Gratton(1989)]{gratton89} 
Gratton, R.\ G.\ 1989, \aap, 208, 171

\bibitem[Gratton \& Sneden(1994)]{gs94}
Gratton, R.\ G., \& Sneden, C.\ 1994, \aap, 287, 927

\bibitem[Griffin(1968)]{griffin68}
Griffin, R.\ F.\ 1968, A Photometric Atlas of the Spectrum of Arcturus $\lambda\lambda$3600--8825\AA\ (Cambridge: Cambridge Philosophical Society)

\bibitem[Hannaford et al.(1985)]{hannaford+85}
Hannaford, P., Lowe, R.\ M., Bi\'emont, E., \& Grevesse, N.\ 1985, \aap, 143, 447

\bibitem[Hannaford et al.(1982)]{hannaford+82}
Hannaford, P., Lowe, R.\ M., Grevesse, N., Bi\'emont, E., \& Whaling, W.\ 1982, \apj, 261, 736 

\bibitem[Hansen et al.(2004)]{hansen+04}
Hansen, B.\ M.\ S., et al.\ 2004, \apjs, 155, 551

\bibitem[Herbig(1975)]{herbig75}
Herbig, G.\ H.\ 1975, \apj, 196, 129

\bibitem[Hill et al.(2002)]{hill+02}
Hill, V., et al.\ 2002, \aap, 387, 560

\bibitem[Hoffman, Woosley, \& Qian(1997)]{hwq97}
Hoffman, R.\ D., Woosley, S.\ E., \& Qian, Y.-Z.\ 1997, \apj, 482, 951

\bibitem[Holweger \& M\"uller(1974)]{hm74}
Holweger, H.\ \& M\"uller, E.\ A.\ 1974, Sol.\ Phys., 39, 19

\bibitem[Honda et al.(2004)]{honda+04}
Honda, S., Aoki, W., Kajino, T., Ando, H., Beers, T.\ C., Izumiura, H., Sadakane, K.\ \& Takada-Hidai, M.\ 2004, \apj, 607, 474

\bibitem[Imura et al.(2003)]{imura+03}
Imura, H., et al.\ 2003, Phys. Rev. C, 68, 054238

\bibitem[Ivans et al.(2001)]{ivans+01}
Ivans, I.\ I., Kraft, R.\ P., Sneden, C., Smith, G.\ H., Rich, M.\ R.\ \& Shetrone, M.\ 2001, \aj, 122, 1438

\bibitem[Ivans et al.(2005)]{ivans+05} 
Ivans, I.\ I., Sneden, C., Gallino, R., Cowan, J.\ J., \& Preston, G.\ W.\ 2005, \apj, 627, L165

\bibitem[Ivans et al.(2003)]{ivans+03} 
Ivans, I.\ I., Sneden, C., James, C.\ R., Preston, G.\ W., Fulbright, J.\ P., H\"oflich, P.\ A., Carney, B.\ W., \& Wheeler, J.\ C.\ 2003, \apj, 592, 906

\bibitem[Ivans et al.(1999)]{ivans+99}
Ivans, I.\ I., Sneden, C., Kraft, R.\ P., Suntzeff, N.\ B., Smith, V.\ V., Langer, G.\ E.\ \& Fulbright, J.\ P.\ 1999, \aj, 118, 1273

\bibitem[Ivarsson et al.(2001)]{ivarsson+01}
Ivarsson, S., Litz\'en, U., \& Wahlgren, G.\ M.\ 2001, \physscr, 64, 455

\bibitem[Ivarsson et al.(2003)]{ivarsson+03}
Ivarsson, S., et al.\ 2003, \aap, 409, 1141

\bibitem[Jimenez et al.(2003)]{jimenez+03}
Jimenez, R., Verde, L., Treu, T.\ \& Stern, S.\ 2003, \apj, 593, 622

\bibitem[Johnson \& Bolte(2001)]{jb01}
Johnson, J.\ A., \& Bolte, M.\ 2001, \apj, 554, 888

\bibitem[Johnson \& Bolte(2004)]{jb04}
Johnson, J.\ A., \& Bolte, M.\ 2004, \apj, 605, 462

\bibitem[Johnson, Ivans \& Stetson(2005)]{jis05}
Johnson, J.\ A., Ivans, I.\ I.\ \& Stetson, P.\ B.\ 2005, \aj, preprint doi:10.1086/498882

\bibitem[K\"appeler, Beer \& Wisshak(1989)]{kbw89}
K\"appeler, F., Beer, H.\ \& Wisshak, K.\ 2989, Rep\. Prog.\ Phys., 52, 945

\bibitem[Klose et al.(2002)]{kfw02}
Klose, J.\ Z., Fuhr, J.\ R., \& Wiese, W.\ L.\ 2002, J.\ Phys.\ Chem.\ Ref.\ Dat., 31, 217

\bibitem[Kohri, Narayan, \& Piran(2005)]{knp05}
Kohri, K., Narayan, R., \& Piran, T.\ 2005, \apj, 629, 341

\bibitem[Kraft \& Ivans(2003)]{ki03}
Kraft, R.\ P., \& Ivans, I.\ I.\ 2003, PASP, 115, 143

\bibitem[Kratz et al.(2004)]{kratz+04}
Kratz, K.-L., Pfeiffer, B., Cowan, J.\ J., \& Sneden, C.\ 2004, New Astronomy Reviews, 48, 105 

\bibitem[Krauss \& Chaboyer(2003)]{kc03}
Krauss, L.\ M.\ \& Chaboyer, B.\ 2003, Science, 299, 65

\bibitem[Kurucz(1998)]{kurucz98}
Kurucz, R.\ L.\ 1998, in Fundamental Stellar Properties: The Interaction between Observation and Theory, IAU Symp. 189, (ed.) T.\ R.\ Bedding, A.\ J.\ Booth and J.\ Davis (Dordrecht: Kluwer), 217

\bibitem[Kurucz \& Bell(1995)]{kb95}
Kurucz, R.\ L.\ \& Bell, B.\ 1995, 1995 Atomic Line Data, Kurucz CD-ROM \#23, Cambridge, MA: Smithsonian Astrophysical Observatory

\bibitem[Kurucz et al.(1984)]{kurucz+84}
Kurucz, R.\ L., Furenlid, I., Brault, J., \& Testerman, L\. 1984, Solar Flux Atlas from 296 to 1300 nm (Cambridge: Harvard Univ. Press) 

\bibitem[Kwiatkowski et al.(1982)]{kwiatkowski+82}
Kwiatkowski, M., Zimmermann, P., Bi\'emont, E., \& Grevesse, N.\ 1982, \aap, 112, 337

\bibitem[Kwiatkowski et al.(1984)]{kwiatkowski+84}
Kwiatkowski, M., Zimmermann, P., Bi\'emont, E., \& Grevesse, N.\ 1984, \aap, 135, 59

\bibitem[Lawler et al.(2001a)]{lawler+01a}
Lawler, J.\ E., Bonvallet, G., \&  Sneden, C.\  2001a, \apj, 556, 452

\bibitem[Lawler et al.(2005)]{lawler+05}
Lawler, J.\ E., Den Hartog, E.\ A., Sneden, C., \& Cowan, J.\ J.\ 2005, \apj, in press

\bibitem[Lawler et al.(2004)]{lawler+04}
Lawler, J.\ E., Sneden, C., \& Cowan, J.\ J.\ 2004, \apj, 604, 850

\bibitem[Lawler et al.(2001b)]{lawler+01b}
Lawler, J.\ E., Wickliffe, M.\ E., Cowley, C.\ R., \&  Sneden, C.\ 2001b, \apjs, 137, 341

\bibitem[Lawler et al.(2001c)]{lawler+01c}
Lawler, J.\ E., Wickliffe, M.\ E., den Hartog, E.\ A., \& Sneden, C.\ 2001c, \apj, 563, 1075

\bibitem[Learner \& Thorne(1988)]{lt88}
Learner, R.\ C.\ M.\ \& Thorne, A.\ P.\ 1988, J.\ Opt.\ Soc.\ Am.\ B, 10, 2045

\bibitem[Li et al.(2001)]{li+01}
Li, G.-W., Zhang, X.-M., Lu, F.-Qu., Peng, X.-J., \& Yang F.-J.\ 2001, Jpn.\ J.\ Appl.\ Phys., 40, 2508

\bibitem[Lodders(2003)]{lodders03} 
Lodders, K.\ 2003, \apj, 591, 1220

\bibitem[Martin et al.(1978)]{martin+78}
Martin, W.\ C., Zalubas, R., \& Hagan, L.\ 1978, Atomic Energy Levels: The Rare Earth Elements, NSRDSNBS 60 (Washington: U.S.G.P.O.), 174

\bibitem[McWilliam(1997)]{mcw97}    
McWilliam, A., 1997, \araa, 35, 503

\bibitem[McWilliam(1998)]{mcw98}    
McWilliam, A., 1998, \aj, 115, 1640

\bibitem[McWilliam et al.(1995)]{mcw+95} 
McWilliam A., Preston, G.\ W., Sneden, C.\ \& Searle, L.\ 1995, \aj, 109, 2757

\bibitem[McWilliam \& Searle(1999)]{ms99}
McWilliam, A.\ \& Searle, L.\ 1999, Ap\&SS, 265, 133

\bibitem[Meyer(2005)]{meyer05}
Meyer, B.\ S.\ 2005, in Chondrites and Protoplanetary Disk (ed.) A.\ N.\ Krot, E.\ R.\ D.\ Scott and B.\ Reipurth, ASP Conf.\ Series 341, 515

\bibitem[Meyer \& Clayton(2000)]{mc00}
Meyer, B.\ S.\ \& Clayton, D.\ C.\ 2000, \ssr, 92, 133

\bibitem[Migdalek \& Baylis(1987)]{mb87}
Migdalek, J., \& Baylis, W.\ E.\ 1987, Can.\ J.\ Phys., 65, 1612

\bibitem[Mishenina \& Kovtyukh(2001)]{mk01}
Mishenina, T.\ V.\ \& Kovtyukh, V.\ V.\ 2001, \aap, 370, 951

\bibitem[Mishenina et al.(2002)]{mishenina+02}
Mishenina, T.\ V., Kovtyukh, V.\ V., Soubiran, C., Travaglio, C., \& Busso, M.\ 2002, \aap, 396, 189

\bibitem[Moore et al.(1966)]{mmh66}
Moore, C.\ E., Minnaert, M.\ G.\ J., \& Houtgast, J.\ 1966, The Solar Spectrum 2934~\AA\ to 8770~\AA, NBS Monograph 61 (Washington: U.S.\ Gov.)

\bibitem[Musaev et al.(1999)]{musaev+99} 
Musaev, F., Galazutdinov, G., Sergeev, A., Karpov, N., \& Pod'yachev, Y.\ 1999, Kinematics Phys.\ Select Bodies, 15, 216

\bibitem[Nakamura et al.(1999)]{nakamura+99}
Nakamura, T., Umeda, H., Nomoto, K., Thielemann, F.-K.\ \& Burrows, A.\ 1999, \apj, 517, 193

\bibitem[Nilsson et al.(1991)]{njk91}
Nilsson, A.\ E., Johansson, S., \& Kurucz, R.\ L.\ 1991, \physscr, 44, 226

\bibitem[Nilsson et al.(2005)]{nilsson+05}
Nilsson, A.\ E., Ljung, L., Lundberg, H., \& Nielsen, K.\ E.\ 2005, \aap, accepted
 
\bibitem[Nilsson et al.(2002)]{nilsson+02}
Nilsson, H., Zhang, Z.\ G., Lundberg, H., Johansson, S., \& Nordstr\"om, B.\ 2002, \aap, 382, 368

\bibitem[O'Brian et al.(1991)]{obrian+91}
O'Brian, T., Wickliffe, M., Lawler, J., Whaling, W.\ \& Brault, J.\ 1991, J.\ Opt.\ Soc.\ Am., B8, 1185

\bibitem[O'Brien et al.(2003)]{obrien+03}
O'Brien, S., Dababneh, S., Heil, M., K\"appeler, F., Plag, R., Reifarth, R., Gallino, R.\ \& Pignatari, M.\ 2003, Phys.\ Rev.\ C, 68, 035801

\bibitem[Otsuki et al.(2000)]{otsuki+00}
Otsuki, T., Tagoshi, H., Kajino, T.\ \& Wanajo, S.\ 2000, \apj, 533, 424

\bibitem[Pagel(1989)]{pagel89}
Pagel, B.\ E.\ J.\ 1989, in Evolutionary phenomena in galaxies, (ed.) J.\ E.\ Beckman \& B.\ E.\ J.\ Pagel, (Cambridge: Cambridge Univ.\ Press), 201

\bibitem[Palmeri et al.(2000)]{palmeri+00}
Palmeri, P., Quinet, P., Wyart, J.-F., \& Bi\'emont, E.\ 2000, \physscr, 61, 323

\bibitem[Persson(1997)]{persson97}
Persson, J.\ R.\ 1997, Z.\ Phys.\ D, 42, 259

\bibitem[Pfeiffer et al.(2001)]{pok01}
Pfeiffer, B., Ott, U., \& Kratz, K.-L.\ 2001, Nucl.\ Phys.\ A, 688, 575

\bibitem[Pfeiffer, Kratz, \& Thielemann(1997)]{pkt97}
Pfeiffer, B., Kratz, K.-L., \& Thielemann, F.-K.\ 1997, Z.\ Phys.\ A,\ 357, 235

\bibitem[Plez et al.(2005)]{plez+05}
Plez, B., Hill, V., Cayrel, R., Spite, M., Barbuy, B., Beers, T.\ C., Bonifacio, P., Primas, F., \&  Nordstr\"om, B.\ 7 2004, \aap, 428, 9

\bibitem[Prochaska \& McWilliam(2000)]{pm00} 
Prochaska, J.\ X.\ \& McWilliam, A.\ 2000, \apj, 537, L57

\bibitem[Qian(2003)]{qian03}
Qian, Y.-Z.\ 2003, Progress in Particle and Nuclear Physics, 50, 153

\bibitem[Qian \& Wasserburg(2000)]{qw00}
Qian, Y.-Z.\ \& Wasserburg, G.\ J.\ 2000, Phys.\ Reports, 333, 77

\bibitem[Qian \& Wasserburg(2001)]{qw01}
Qian, Y.-Z.\ \& Wasserburg, G.\ J.\ 2001, \apj, 552, L55

\bibitem[Qian \& Wasserburg(2003)]{qw03}
Qian, Y.-Z.\ \& Wasserburg, G.\ J.\ 2003, \apj, 588, 1099

\bibitem[Ram\'irez \& Mel\'endez(2005)]{rm05} 
Ram\'irez, I.\ \& Mel\'endez, J.\ 2005, \apj, 626, 465

\bibitem[Rosman \& Taylor(1997)]{rt97}
Rosman, K.\ J.\ R., \& Taylor, P.\ D.\ P.\ 1998, J.\ Phys.\ Chem.\ Ref.\ Data, 27, 1275

\bibitem[Ross \& Aller(1972)]{ra72}
Ross, J.\ E., \& Aller, L.\ H.\ 1972, \solphys, 25, 30

\bibitem[Rosswog et al.(1999)]{rosswog+99} 
Rosswog, S., Liebend\"orfer, M., Thielemann, F.-K., Davies, M.\ B., Benz, W.\ \& Piran, T.\ 1999, \aap, 341, 499

\bibitem[Russell et al.(1941)]{russell+41}
Russell, H.\ N., Albertson, W., \& Davis, D.\ N.\ 1941, Phys.\ Rev., 60, 641

\bibitem[Ryan et al.(1996)]{rnb96} 
Ryan, S.\ G., Norris, J.\ E.\ \& Beers, T.\ C.\ 1996, \apj, 471, 254

\bibitem[Sasaqui et al.(2005)]{skb05}
Sasaqui, T., Kajino, T.\ \& Balantekin, A.\ B.\ 2005, \apj, submitted

\bibitem[Schatz et al.(2002)]{schatz+02}
Schatz, H.,  Toenjes, R.,  Kratz, K-L., Pfeiffer, B.,  Beers, T.\ C., Cowan, J.\ J., \& Hill, V.\ 2002 \apj, 579, 626  

\bibitem[Schlegel, Finkbeiner, \& Davis(1998)]{sfd98} 
Schlegel, D.\ J., Finkbeiner, D.\ P., \& Davis, M.\ 1998, \apj, 500, 525

\bibitem[Schlegel(2003)]{schlegel03}
Schlegel, E.\ M.\ 2003, \aj, 125, 1426

\bibitem[Simmerer et al.(2004)]{jensim+04} 
Simmerer, J., Sneden, C., Cowan, J.\ J., Collier, J., Woolf, V.\ M., Lawler, J.\ E.\ 2004, \apj, 617, 1091

\bibitem[Simmerer et al.(2003)]{jensim+03} 
Simmerer, J., Sneden, C., Ivans, I.\ I., Kraft, R.\ P., Shetrone, M.\ D., \& Smith, V.\ V.\ 2003, \aj, 125, 2018

\bibitem[Sneden(1973)]{sneden73} 
Sneden, C.\ 1973, \apj, 184, 839

\bibitem[Sneden \& Cowan(2003)]{sc03}
Sneden, C., \& Cowan, J.\ J.\ 2003, Science, 299, 70

\bibitem[Sneden et al.(1998)]{sneden+98}
Sneden, C., Cowan, J.\ J., Burris, D.\ L., \& Truran, J.\ W.\ 1998, \apj, 496, 235

\bibitem[Sneden et al.(2000a)]{sneden+00a}
Sneden, C., Cowan, J.\ J., Ivans, I.\ I., Fuller, G.\ M., Burles, S., Beers, T.\ C., \& Lawler, J.\ E.\  2000a, \apj, 533, L139

\bibitem[Sneden et al.(2002)]{sneden+02} 
Sneden, C., Cowan, J.\ J., Lawler, J.\ E., Burles, S., Beers, T.\ C., Fuller, G.\ M.\ 2002, \apj, 566, L25

\bibitem[Sneden et al.(2000b)]{sneden+00b}
Sneden, C., Johnson, J., Kraft, R.\ P., Smith, G.\ H., Cowan, J.\ J., \& Bolte, M.\ S.\ 2000b, \apj, 536, L85

\bibitem[Sneden et al.(2004)]{sneden+04} 
Sneden, C., Kraft, R.\ P., Guhathakurta, P., Peterson, R.\ C., \& Fulbright, J.\ P.\ 2004, \aj, 127, 2162

\bibitem[Sneden et al.(1991b)]{sneden+91b} 
Sneden, C., Kraft, R.\ P., Prosser, C.\ F.\ \& Langer, G.\ E.\ 1991b, \aj, 102, 2001

\bibitem[Sneden et al.(1996)]{sneden+96}
Sneden, C., McWilliam, A., Preston, G.\ W., Cowan, J.\ J., Burris, D.\ L., \& Armosky, B.\ J.\ 1996, \apj, 467, 819

\bibitem[Sneden, Pilachowski, \& VandenBerg(1986)]{spv86}
Sneden, C., Pilachowski, C.\ A.\ \& VandenBerg, D.\ A.\ 1986, \apj, 311, 826

\bibitem[Sneden et al.(2003)]{sneden+03} 
Sneden, C., et al.\ 2003, \apj, 591, 936

\bibitem[Soubiran, Katz \& Cayrel(1998)]{skc98} 
Soubiran, C., Katz, D., \& Cayrel, R.\ 1998, \aap, 133, 221

\bibitem[Spergel et al.(2003)]{spergel+03}
Spergel, D.\ N., et al.\ 2003, \apjs, 148, 175

\bibitem[Tegmark et al.(2004)]{tegmark+04}
Tegmark, M., et al.\ 2004, Phys.\ Rev.\ D, 69, 103501

\bibitem[Terasawa et al.(2002)]{terasawa+02} 
Terasawa, M., Sumiyoshi, K., Yamada, S., Suzuki, H.\ \& Kajino, T.\ 2002, \apj, 578, 137

\bibitem[Thielemann et al.(2001)]{thielemann+01}
Thielemann, F.-K., et al.\ 2001, in 27th International Cosmic Ray Conference. Invited, Rapporteur, and Highlight Papers. 07-15 August, 2001. Hamburg, Germany.\ (ed.) R.\ Schlickeiser, Under the auspices of the International Union of Pure and Applied Physics (IUPAP), 52

\bibitem[Travaglio et al.(2004)]{travaglio+04}
Travaglio, C., Gallino, R., Arnone, R., Cowan, J.\ J., Jordan, F.\ \& Sneden, C.\ 2004, \apj, 601, 864

\bibitem[Truran et al.(2001)]{tcf01}
Truran, J.\ W., Cowan, J.\ J., \& Fields, B.\ D.\ 2001, Nucl.\ Phys.\ A, 688, 330

\bibitem[Truran et al.(2002)]{truran+02}
Truran, J.\ W., Cowan, J.\ J., Pilachowski, C.\ A., \& Sneden, C.\ 2002, \pasp,  114, 1293

\bibitem[Tull et al.(1995)]{tull+95} 
Tull, R.\ G., MacQueen, P.\ J., Sneden, C., \& Lambert, D.\ L.\ 1995, \pasp, 107, 251

\bibitem[Venn et al.(2004)]{venn+04}
Venn, K.\ A., Irwin, M., Shetrone, M.\ D., Tout, C.\ A., Hill, V.\ \& Tolstoy, E.\ 2004, \aj, 128

\bibitem[Villemoes et al.(1992)]{villemoes+92}
Villemoes, P., Arnesen, A., Heijkenskj\"old, F., Kastberg, A., \& Larsson, M.\ O.\ 1992, \physscr, 46, 45

\bibitem[Voges et al.(1999)]{voges+99}
Voges, W., et al.\ 1999, \aap, 349, 389

\bibitem[Vogt et al.(1994)]{vogt+94} 
Vogt, S.\ S., et al.\ 1994, SPIE, 2198, 362

\bibitem[Wallerstein et al.(1963)]{wall+63} 
Wallerstein, G., Greenstein, J.\ L., Parker, R., Helfer, H.\ L., \& Aller, L.\ H.\ 1963, \apj, 137, 280

\bibitem[Wanajo et al.(2002)]{wanajo+02}
Wanajo, S., Itoh, N., Ishimaru, Y., Nozawa, S.\ \& Beers, T.\ C.\ 2002, \apj, 577, 853

\bibitem[W\"annstr\"om et al.(1994)]{wannstrom+94}
W\"annstr\"om, A., Gough, D.\ S., \& Hannaford, P.\ 1994, Z.\ Phys.\ D, 29, 39

\bibitem[Wasserburg et al.(1996)]{wbg96}
Wasserburg, G.\ J., Busso, M.\ \& Gallino, R.\ 1996, \apj, 466, L109

\bibitem[Wasserburg et al.(2006)]{wasserburg+06}
Wasserburg, G.\ J., Busso, M., Gallino, R.,\ \& Nollett, K.\ M.\ 2006, Nucl.\ Phys.\ A, in press

\bibitem[Wasserburg \& Qian(2000)]{wq00}
Wasserburg, G.\ J.\ \& Qian, Y.-Z.\ 2000, \apj, 529, L21

\bibitem[Westin et al.(2000)]{westin+00}
Westin, J., Sneden, C., Gustafsson, B.\ \& Cowan, J.\ J.\ 2000, \apj, 530, 783

\bibitem[Whaling et al.(2002)]{whaling+02}
Whaling, W., Anderson, W.\ H.\ C., Carle, M.\ T., Brault, J.\ W., \& Zarem, H.\ A.\ 2002, J.\ Res.\ Natl.\ Inst.\ Stand.\ \& Tech., 107, 149

\bibitem[Whaling \& Brault(1988)]{wb88}
Whaling, W., \& Brault, J.\ W.\ 1988, \physscr, 38, 707

\bibitem[Wheeler et al(1998)]{wch98}
Wheeler, J.\ C.\, Cowan, J.\ J.\ \& Hillebrandt, W.\ 1998, \apj, 493, L101

\bibitem[Wickliffe \& Lawler(1985)]{wl85}
Wickliffe, M.\ E., \& Lawler, J.\ E.\ 1997, J.\ Opt.\ Soc.\ Am.\ B, 14, 737

\bibitem[Wickliffe et al.(2000)]{wickliffe+00}
Wickliffe, M.\ E., Lawler, J.\ E., \& Nave, G.\ 2000, J.\ Quant.\ Spec.\ Rad.\
Trans., 66, 363

\bibitem[Wickliffe et al.(1994)]{wickliffe+94}
Wickliffe, M.\ E., Salih, S., \& Lawler, J.\ E.\ 1994, J.\ Quant.\ Spec.\  Rad.\ Trans., 51, 545

\bibitem[Woosley \& Hoffman(1992)]{wh92}
Woosley, S.\ E.\ \& Hoffman, R.\ D.\ 1992, \apj, 395, 202

\bibitem[Xu et al.(2003)]{xu+03}
Xu, H., Jiang, Z., Zhang, Z., Dai, Z., Svanberg, S., Quinet, P.\ \& Bi\'emont, E.\ 2003, J.\ Phys.\ B, 36, 1771

\bibitem[Yushchenko et al.(2002)]{yu+02} 
Yushchenko, A., et al.\ 2002, JKAS, 35, 209

\bibitem[Yushchenko et al.(2005)]{yu+05} 
Yushchenko, A., et al.\ 2005, \aap, 430, 255

\end{thebibliography}
\end{document}